\documentclass[aps,prb,twocolumn,showpacs,showkeys,amssymb]{revtex4}

\usepackage{graphics}% Include figure files
\usepackage{bm}% bold math
\usepackage{amsmath}
\usepackage{subfigure}

\newcommand{\wt}{\widetilde}
\newcommand{\be}{\begin{equation}}
\newcommand{\ee}{\end{equation}}
\newcommand{\bea}{\begin{eqnarray}}
\newcommand{\eea}{\end{eqnarray}}

\newcommand{\vect}[1]{\mathbf{#1}}
\newcommand{\GG}{\mathcal{G}}
\newcommand{\G}{\widetilde{\mathcal{G}}} % for fancy G in green function

\newcommand{\Gzero}{\mathcal{G}^{(0)}}
\newcommand{\Sig}{\widetilde{\Sigma}} % for self-energy
\newcommand{\parg}{\left(p_1,p_2\right)}

\begin{document}
\title{Low temperature thermal conductivity in a $d$-wave superconductor with
coexisting charge order: Effect of self-consistent disorder and
vertex corrections}

\author{Philip R. Schiff}
\author{Adam C. Durst}
   \affiliation{Department of Physics and Astronomy, Stony Brook University, Stony Brook, NY 11794-3800, USA}
    \email{pschiff@grad.physics.sunysb.edu}
    \email{adam.durst@stonybrook.edu}
\date{November 1, 2008}

\begin{abstract}
Given the experimental evidence of charge order in the underdoped
cuprate superconductors, we consider the effect of coexisting
charge order on low-temperature thermal transport in a $d$-wave
superconductor. Using a phenomenological Hamiltonian that
describes a two-dimensional system in the presence of a ${\bf
Q}=(\pi,0)$ charge density wave and $d$-wave superconducting
order, and including the effects of weak impurity scattering, we
compute the self-energy of the quasiparticles within the
self-consistent Born approximation, and calculate the
zero-temperature thermal conductivity using linear response
formalism. We find that vertex corrections within the ladder
approximation do not significantly modify the bare-bubble result
that was previously calculated. However, self-consistent treatment
of the disorder does modify the charge-order-dependence of the
thermal conductivity tensor, in that the magnitude of charge order
required for the system to become effectively gapped is
renormalized, generally to a smaller value.
\end{abstract}

\maketitle

%%%%%%%%%%%%%%%%%%%%%%%%%%%%%%%%%%%%%%%%%%%%%%%%%%%%%%%%%%%%%%%%%%%%%%%%%%%%%%%%%%%%%%
\section{Introduction}
%%%%%%%%%%%%%%%%%%%%%%%%%%%%%%%%%%%%%%%%%%%%%%%%%%%%%%%%%%%%%%%%%%%%%%%%%%%%%%%%%%%%%%
The superconducting phase of the cuprate superconductors exhibits
$d$-wave pairing symmetry.\cite{har01} As such, there exist four
nodal points on the two-dimensional Fermi surface at which the
quasiparticle excitations are gapless, and quasiparticles excited
in the vicinity of a node behave like massless Dirac
fermions.\cite{lee02,alt01,ore01}. The presence of impurities
enhances the density of states at low energy\cite{gor01} resulting
in a universal limit $(T\rightarrow0,~\Omega\rightarrow0)$ where
the thermal conductivity is independent of
disorder.\cite{lee01,hir01,hir02,hir03,graf01,sen01,dur01}
Calculations have shown that the thermal conductivity retains this
universal character even upon the inclusion of vertex
corrections.\cite{dur01} Experiments have confirmed the validity
of this quasiparticle picture of transport by observing their
universal-limit contribution to the thermal conductivity, and
thereby measuring the anisotropy of the the Dirac nodes,
$v_f/v_\Delta$.\cite{tai01,chi01,chi02,nak01,pro01,sut01,hil01,sun01,sut02,haw01,sun02}

For some time, there has been significant interest
\cite{kiv01,pod01,li01,che01,seo01} in the idea of additional
types of order coexisting with $d$-wave superconductivity (dSC) in
the cuprates. And in recent years, as the underdoped regime of the
phase diagram has been explored in greater detail, evidence of
coexisting order has grown substantially \cite{kiv01}.
Particularly intriguing has been the evidence of checkerboard
charge order revealed via scanning tunnelling microscopy (STM)
experiments.
\cite{hof01,hof02,how01,ver01,mce01,han01,mis01,mce02,koh01,boy01,han02,pas01,wis01,koh02}

And if charge order coexists with $d$-wave superconductivity in
the underdoped cuprates, it begs the question of how the
quasiparticle excitation spectrum is modified.  Previous work
\cite{ber01} has shown that even with the addition of a charge or
spin density wave to the dSC hamiltonian, the low-energy
excitation spectrum remains gapless as long as a harmonic of the
ordering vector does not nest the nodal points of the combined
hamiltonian.  However, if the coexisting order is strong enough,
the nodal points can move to $k$-space locations where they are
nested by the ordering vector, at which point the excitation
spectrum becomes fully gapped. \cite{par01,gra01,voj01}

Such a nodal transition should have dramatic consequences for
low-temperature thermal transport, the details of which were
studied in Ref.~\onlinecite{dur02}.  That paper considered the
case of a conventional $s$-wave charge density wave (CDW) of wave
vector $\vect{Q}=(\pi,0)$ coexisting with $d$-wave
superconductivity.  It showed that the zero-temperature thermal
conductivity vanishes, as expected, once charge order is of
sufficient magnitude to gap the quasiparticle spectrum.  In
addition, the dependence of zero-temperature thermal transport was
calculated and revealed to be disorder-dependent.  Hence, in the
presence of charge order, the universal-limit is no longer
universal.  This result is in line with the results of recent
measurements \cite{hus01,tak01,sun05,and01,sun03,sun04,haw02} of
the underdoped cuprates, as well as other calculations
\cite{gus01,ander01}.

We extend the work of Ref.~\onlinecite{dur02} herein.  We consider
the same physical system, but employ a more sophisticated model of
disorder that includes the effects of impurity scattering within
the self-consistent Born approximation.  We find that this
self-consistent model of disorder requires that off-diagonal
components be retained in our matrix self-energy.  These
additional components lead to a renormalization of the critical
value of charge order beyond which the thermal conductivity
vanishes.  Furthermore, we include the contribution of vertex
corrections within our diagrammatic thermal transport calculation.
While vertex corrections become more important as charge order
increases, especially for long-ranged impurity potentials, we find
that for reasonable parameter values, they do not significantly
modify the bare-bubble result.

In Sec.~\ref{sec:model}, we introduce the model hamiltonian of the
dSC+CDW system, describe the effect charge ordering has on the
nodal excitations, and present our model for disorder.  In
Sec.~\ref{sec:SCBA}, a numerical procedure for computing the
self-energy within the self-consistent Born approximation is
outlined.  The results of its application in the relevant region
of parameter space are presented in Sec.~\ref{sec:SCBAresults}. In
Sec.~\ref{sec:thermalconductivity}, we calculate the thermal
conductivity using a diagrammatic Kubo formula approach, including
vertex corrections within the ladder approximation. An analysis of
the vertex-corrected results and a calculation of the clean-limit
thermal conductivity is presented in Sec.~\ref{sec:analysis}. Also
in this section, we discuss how our self-consistent model of
disorder renormalizes the nodal transition point, the value of
charge order parameter at which the nodes effectively vanish.
Conclusions are presented in Sec.~\ref{sec:conc}.

%%%%%%%%%%%%%%%%%%%%%%%%%%%%%%%%%%%%%%%%%%%%%%%%%%%%%%%%%%%%%%%%%%%%%%%%%%%%%%%%%%%%%%%%
\section{Model}
\label{sec:model}
\begin{figure}
  \centerline{\resizebox{3.25 in}{!}{\includegraphics{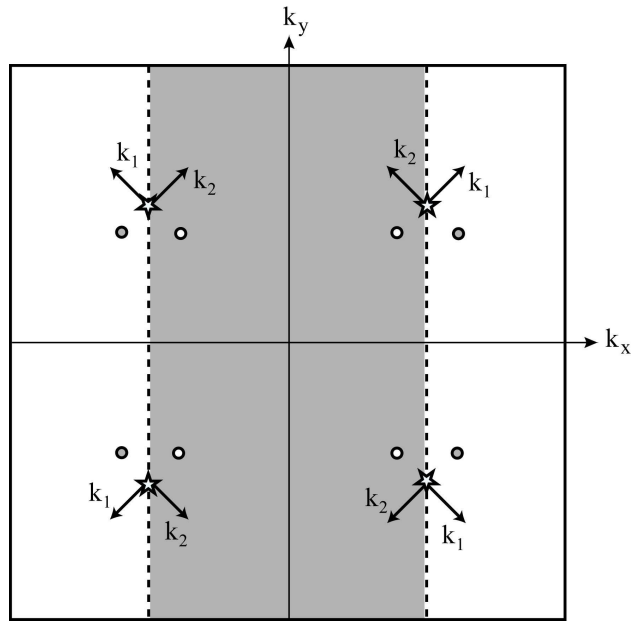}}}
  \caption{Illustrated is the Brillouin zone for our model, reduced to the
  shaded region by unit-cell-doubling charge order. The $\psi=0$ nodal locations
  are illustrated by white dots. They are displaced by a distance $k_0$ from the
  $(\pm\frac{\pi}{2},\pm\frac{\pi}{2})$ points (stars). As the charge density
  wave's amplitude increases, the location of the gapless excitations evolves
  along curved paths toward the $(\pm\frac{\pi}{2},\pm\frac{\pi}{2})$ points,
  until $\psi$ reaches $\psi_c$, when the spectrum becomes gapped because the
  nodes are nested by the charge density wave-vector. The gray dots depict the
  images of the nodes in the second reduced Brillouin zone.}
  \label{fig:BZfig}
\end{figure}

%%%%%%%%%%%%%%%%%%%%%%%%%%%%%%%%%%%%%%%%%%%%%%%%%%%%%%%%%%%%%%%%%%%%%%%%%%%%%%%%%%%%%%%%
We employ the phenomenological hamiltonian of
Ref.~\onlinecite{dur02} in order to calculate the low-temperature
thermal conductivity of the fermionic excitations of a $d$-wave
superconductor with a $\vect{Q}=(\pi,0)$ charge density wave, in
the presence of a small but nonzero density of point-like impurity
scatterers. The presence of $d$-wave superconducting order
contributes a term to the hamiltonian \be
 H_{dSC}=\frac{1}{2}\sum_{k\alpha}\Big(\epsilon_k c_{k\alpha}^\dagger
 c_{k\alpha}+\Delta_k c_{k\alpha}^\dagger c_{-k\beta}^\dagger\Big)+\mathrm{h.c.}
\ee where $\epsilon_k$ is a typical tight-binding dispersion, and
$\Delta_k$ an order parameter of $d_{x^2-y^2}$ symmetry. Due to
the $d$-wave nature of the gap, nodal excitations exist in the
$(\pm\pi,\pm\pi)$ directions with respect to the origin. The
locations of these nodes in the absence of charge ordering are
close to the points $(\pm \pi/2,\pm \pi/2)$, and are denoted with
white dots in Fig.~\ref{fig:BZfig}. These low energy excitations
are massless anisotropic Dirac fermions. That is, the electron
dispersion and pair function are linear functions of momentum in
the vicinity of these nodal locations. We will refer to the slopes
of the electron dispersion and pair function, defined by
$\vect{v}_f\equiv\frac{\partial\epsilon_k}{\partial\bf{k}}$ and
$\vect{v}_\Delta\equiv\frac{\partial\Delta_k}{\partial\bf{k}}$, as
the Fermi velocity and gap velocity respectively. The energy of
the quasiparticles in the vicinity of the nodes is given by
$E_k=\sqrt{v_f^2 k_1^2+v_\Delta^2 k_2^2}$, where $k_1$ and $k_2$
are the momentum displacements (from the nodes) in directions
perpendicular to and parallel to the Fermi surface. The
universal-limit $(T\rightarrow 0, \Omega\rightarrow 0)$ transport
properties of these quasiparticles was explored in
Ref.~\onlinecite{dur01}.

While experiments have revealed evidence of a number of varieties
of spin and charge order, the system described in this paper will
be restricted to the addition of a site-centered charge density
wave of wave vector $\vect{Q}=(\pi,0)$, which contributes a term
to the hamiltonian
\be
 H_{CDW}=\sum_{k\alpha}a_k c_{k\alpha}^\dagger c_{k+Q \alpha}+\mathrm{h.c.}
\ee
The charge density wave doubles the unit cell, reducing the
Brillouin zone to the shaded portion seen in Fig.~\ref{fig:BZfig}.
Restricting summations over momentum space to the reduced
Brillouin zone, and invoking the charge density wave's
time-reversal symmetry and commensurability with the reciprocal
lattice, we are able to write the hamiltonian as
\bea
 H&=&\sum_{k}\Psi^\dagger_kH_k \Psi \hspace{10pt} H_k=H^{dSC}_k+H^{CDW}_k,
\eea
where
\be
 H_k=
 \begin{pmatrix}
  \epsilon_k & \Delta_k & \psi & 0 \\
  \Delta_k & -\epsilon_k & 0 & -\psi\\
  \psi & 0 & \epsilon_{k+Q} & \Delta_{k+Q} \\
  0 & -\psi & \Delta_{k+Q} & -\epsilon_{k+Q}
 \end{pmatrix},
\ee
is a matrix in the basis of extended-Nambu vectors,
\be
 \Psi_k=
 \begin{pmatrix}
   c_{k\uparrow} \\ c^\dagger_{-k\downarrow} \\ c_{k+Q\uparrow} \\ c^\dagger_{-k-Q\downarrow}
 \end{pmatrix}
 \hspace{10pt}
 \Psi^\dagger_k=
 \begin{pmatrix}
  c^\dagger_{k\uparrow} & c_{-k\downarrow} & c^\dagger_{k+Q\uparrow} & c_{-k-Q\downarrow}
 \end{pmatrix}
\ee and $\psi$ represents the constant value taken at the nodes by
the charge density wave order parameter $A_k=a_k+a_{k+Q}^*$.

The onset of the charge order modifies the energy spectrum of the
clean hamiltonian so that the locations of the nodes evolve along
curved paths towards the $(\pm\frac{\pi}{2},\pm\frac{\pi}{2})$
points at the edges of the reduced Brillouin zone, as was noted in
Ref.~\onlinecite{par01}. ``Ghost'' nodes, their images in what is
now the second reduced Brillouin zone, evolve the same way, until
the charge density wave is strong enough that the nodes and ghost
nodes collide at those $(\pm\pi/2,\pm\pi/2)$ points. When that
occurs, $\vect{Q}$ nests two of the nodes, gapping the spectrum so
that low temperature quasiparticle transport is no longer
possible. We define the value of $\psi$ at which this occurs as
$\psi_c$. Due to the nodal properties of the quasiparticles, all
functions of momentum space $\vect{k}$ can be parametrized in
terms of a node index $j$, and local coordinates $p_1$ and $p_2$
in the vicinity of each node. We choose to parametrize our
functions using symmetrized coordinates centered at $(\pm \pi/2,
\pm \pi/2)$,
\bea
 \label{eq:parametrization}
 \epsilon_k&=&\psi_c+\beta p_1 \hspace{40pt} \Delta_k=\frac{1}{\beta}p_2 \nonumber\\
 \epsilon_{k+Q}&=&\psi_c+\beta p_2  \hspace{40pt} \Delta_{k+Q}=\frac{1}{\beta}p_1
\eea where we have rescaled $\sqrt{v_f v_\Delta} k_1= p_1$ for the
coordinate normal to Fermi surface, $\sqrt{v_f v_{\Delta}} k_2 =
p_2$ for the coordinate parallel to Fermi surface, and introduced
the definition $\beta\equiv \sqrt{\frac{v_f}{v_{\Delta}}}$. In
this coordinate system, the displacement of the original node
locations from the collision points is given by $\psi_c$. A sum
over momentum space is therefore performed by summing over nodes,
and integrating over each node's contribution, as follows.
\bea
 \label{eq:intparametrization}
 \sum_k f(\vect{k})\rightarrow\frac{1}{2}\sum_{j=1}^4\int\frac{\mathrm{d}^2p}{4\pi^2 v_fv_{\Delta}}f^{(j)}(p_1,p_2)\nonumber\\
 =\frac{1}{8\pi^2 v_fv_{\Delta}}\sum_{j=1}^4\int_{-p_0}^{p_0}\mathrm{d}p_1\int_{-p_0}^{p_0}\mathrm{d}p_2 \,\,f^{(j)}(p_1,p_2)
\eea where the factor of $\frac{1}{2}$ comes from extending the
integrals to all $p_1$ and $p_2$, rather than just the shaded part
depicted in Fig.~\ref{fig:BZfig}, and $p_0$ is a high-energy
cutoff.

At sufficiently low temperatures, the thermal conductivity is
dominated by the nodal excitations, since phonon modes are frozen
out, and other quasiparticles are exponentially rare. Using this
fact, we can calculate the low temperature thermal conductivity of
the system using linear response formalism.

We incorporate disorder into the model by including scattering
events from randomly distributed impurities. Because the
quasiparticles are nodal, only limited information about the
scattering potential is needed, in particular, the amplitudes
$V_1,V_2$ and $V_3$, for intra-node, adjacent node, and opposite
node scattering respectively, as explained in
Ref.~\onlinecite{dur01}. We calculate the thermal conductivity
using linear response formalism, wherein we obtain the retarded
current-current correlation function by analytic continuation of
the corresponding Matsubara correlator \cite{mah01,Fetter}.

In Ref.~\onlinecite{dur02}, using a simplified model for disorder,
where the self-energy was assumed to be a negative imaginary
scalar, the thermal conductivity was calculated as a function of
$\psi$, and found to vanish for $\psi>\psi_c$. We now improve upon
that result by calculating the self-energy within the
self-consistent Born approximation, and by including vertex
corrections within the ladder approximation in our calculation of
the thermal conductivity.

%%%%%%%%%%%%%%%%%%%%%%%%%%%%%%%%%%%%%%%%%%%%%%%%%%%%%%%%%%%%%%%%%%%%%%%%%%%%%%%%%%%%%%%%%%%%%%%%%%%%%%%%%%%%%%%%%%%%%%
\section{Self-Energy}
%%%%%%%%%%%%%%%%%%%%%%%%%%%%%%%%%%%%%%%%%%%%%%%%%%%%%%%%%%%%%%%%%%%%%%%%%%%%%%%%%%%%%%%%%%%%%%%%%%%%%%%%%%%%%%%%%%%%%%
%%%%%%%%%%%%%%%%%%%%%%%%%%%%%%%%%%%%%%%%%%%%%%%%%%%%%%%%%%%%%%%%%%%%%%%%%%%%%%%%%%%%%%%%%%%%%%%%%%%%%%%%%%%%%%%%%%%%%%
\subsection{SCBA Calculation}
\label{sec:SCBA}
\begin{figure}
   \centerline{\resizebox{2.25 in}{!}{\includegraphics{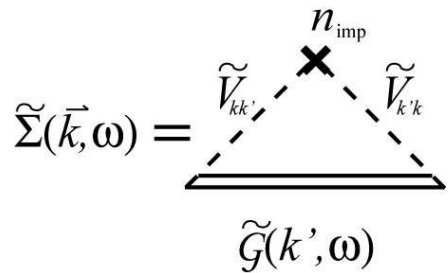}}}
  \caption{Feynman diagram depicting self-energy in the self-consistent
Born approximation. The double line represents the dressed
propagator, the dashed line represents the interaction with the
impurity, and the cross represents the impurity density.}
  \label{fig:bornFeynman}
\end{figure}

%%%%%%%%%%%%%%%%%%%%%%%%%%%%%%%%%%%%%%%%%%%%%%%%%%%%%%%%%%%%%%%%%%%%%%%%%%%%%%%%%%%%%%%%%%%%%%%%%%%%%%%%%%%%%%%%%%%%%%
Within the self-consistent Born approximation (SCBA), the
self-energy tensor is given by
\be
 \label{eq:sigma}
    \Sig(\vect{k},\omega)=n_{\mathrm{imp}}\sum_{k'}\left|V_{kk'}\right|^2
    (\widetilde{
    \sigma_0\otimes\tau_3}
    )
    \G(\vect{k},\omega)
    (\widetilde{
    \sigma_0\otimes\tau_3}
    )
\ee where $n_{\mathrm{imp}}$ is the impurity density and
$\wt{V}_{kk'}=V_{kk'}(\wt{\sigma_0\otimes\tau_3})$ accompanies
each scattering event, as seen in Fig.~\ref{fig:bornFeynman}. The
tilde signifies an operator in the extended-Nambu basis, and the
$\sigma$'s and $\tau$'s are Pauli matrices in charge-order-coupled
and particle-hole spaces respectively.
$\G(\vect{k},\omega)$ is the full Green's function, whose relation
to the bare Green's function $\G_0(\vect{k},\omega)$ and the
self-energy $\Sig(\vect{k},\omega)$ is given by Dyson's equation
\be
 \label{eq:Dyson}
 \G(\vect{k},\omega) = (\G^{-1}_0(\vect{k},\omega)-\Sig(\vect{k},\omega))^{-1},
\ee
 the bare Green's function having been determined by
\be
 \G_0(\vect{k},\omega)=(\omega\wt{\openone}-\wt{H}_k)^{-1}.
\ee Eq.~(\ref{eq:sigma}) and Eq.~(\ref{eq:Dyson}) define a set of
integral equations for the self-energy $\Sig(\vect{k},\omega)$.
For the calculation of the universal-limit thermal conductivity,
it is sufficient to find the zero-frequency limit of the
self-energy. In its present form, $\Sig$ has 32 real components.
Below, we demonstrate that this number can be reduced further to
six components.

If we write the Green's function as
\be
 \G(\vect{k},\omega)=\frac{1}{\GG_{den}}
 \left(
    \begin{array}{cc}
    \GG_A & \GG_B \\
    \GG_C & \GG_D
    \end{array}
 \right),
\ee
where
\be
 \GG_\alpha=\sum_{i=0}^3 \GG_{\alpha i}\tau_i
\ee
then the self-energy can be written as the set of 16 complex
equations (for  $\alpha=\{A,B,C,D\}$, $i=\{0,1,2,3\}$) \bea
 \label{eq:Self}
 \Sigma_{\alpha i}&=&n_{\mathrm{imp}} \sum_{k'}\left|V_{kk'}\right|^2\frac{\xi_i}{\GG_{den}}\GG_{\alpha i}\nonumber\\
 &=&\xi_i c\int\mathrm{d}^2p \frac{\GG_{\alpha i}\parg}{\GG_{den}\parg}
\eea where $ \xi_i=\left\{\begin{array}{c} + 1, \,i=0,3\\-1,\,
i=1,2
\end{array}\right\}$, $c=\frac{n_i(V_1^2+2V_2^2+V_3^2)}{8\pi^2v_fv_\Delta}$,
and the final line is realized by using the notation of
Eq.~(\ref{eq:parametrization}) and
Eq.~(\ref{eq:intparametrization}) and completing the sum over
nodes. From the symmetries of the hamiltonian, we are able to
ascertain certain symmetries the bare Green's function will obey,
specifically, \bea
 \label{eq:greensymmetries}
 \Gzero_{A0}(p_2,p_1)&=&\Gzero_{D0}\parg\\
 \Gzero_{A1}(p_2,p_1)&=&\Gzero_{D1}\parg\nonumber\\
 \Gzero_{A3}(p_2,p_1)&=&\Gzero_{D3}\parg\nonumber\\
 \Gzero_{B0}(p_2,p_1)&=&\Gzero_{C0}\parg\nonumber\\
 \Gzero_{B1}(p_2,p_1)&=&\Gzero_{C1}\parg\nonumber\\
 \Gzero_{B2}(p_2,p_1)&=&\Gzero_{C2}\parg\nonumber\\
 \Gzero_{B3}(p_2,p_1)&=&\Gzero_{C3}\parg\nonumber\\
 \Gzero_{\mathrm{den}}(p_2,p_1)&=&\Gzero_{\mathrm{den}}(p_1,p_2)\nonumber
\eea In addition, the realization that the integration is also
symmetric with respect to exchange of $p_1$ and $p_2$, coupled
with these symmetries, lead to relations for self-energy
components \bea
 \Sigma_{Ai}&=&\Sigma_{Di}\\
 \Sigma_{Bi}&=&\Sigma_{Ci} \hspace{40pt} i=0,1,2,3\nonumber\\
 \Sigma_{B2}&=&\Sigma_{C2}=0\nonumber
\eea so that we see a reduction from 32 components of the
self-energy to 6 independent components:$\{\Sigma_{\alpha
i}\}\equiv\{\Sigma_{A0},\Sigma_{A1},\Sigma_{A3},\Sigma_{B0},\Sigma_{B1},\Sigma_{B3}\}$.
A self-consistent self-energy must therefore satisfy 6 coupled
integral equations given by Eq.~(\ref{eq:Self}).

The self-consistent calculation of the self-energy proceeds by
applying the following scheme: First, a guess is made as to which
self-energy components will be included.  The full Green's
function corresponding to such a self-energy is then obtained from
Dyson's equation, Eq.~(\ref{eq:Dyson}). The quantitative values of
the $\Sigma_{\alpha i}$'s are then determined as follows: An
initial guess for the quantitative values of each of the
$\Sigma_{\alpha i}$'s is made, and the six integrals of
Eq.~(\ref{eq:Self}) are computed numerically, which provides the
next set of guesses for $\{\Sigma_{\alpha i}\}$. This process is
repeated until a stable solution is reached. Finally, the
resulting solutions must be checked that they are consistent with
the initial guess for the form of $\Sig$. If they are, the
self-consistent calculation is complete.

\begin{figure}
  \centerline{\resizebox{3.25 in}{!}{\includegraphics{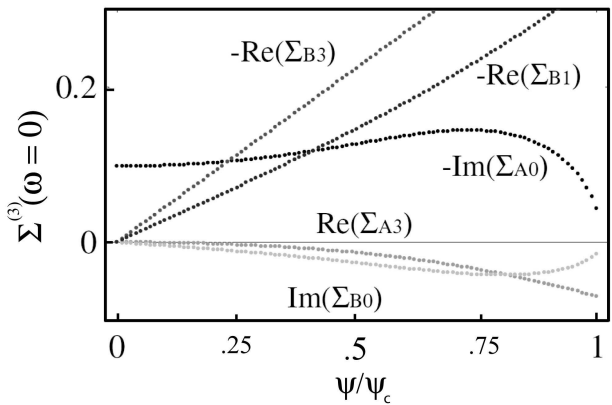}}}
  \caption{Components of self-energy computed using iterative procedure
  described in Sec.~\ref{sec:SCBA}. The third iteration self-energy, $\Sig^{(3)}$,
   is shown here. The dominance of
   $\Gamma_0=-\mathrm{Im}(\Sigma_{A0}),B_1=-\mathrm{Re}(\Sigma_{B1})$,
   and $B_3=-\mathrm{Re}(\Sigma_{B3})$ over other components
   establishes this third iteration as yielding the (approximately) self-consistent value
   of the self-energy.
   $\Sigma_{A1}$ and $\Sigma_{A3}$ overlap.}
  \label{fig:6sigmasBW}
\end{figure}

We begin with the simplest assumption, that $\Sig^{(1)}(\omega) =
-i\Gamma_0 \widetilde{(\sigma_0\otimes\tau_0)}$, where $\Gamma_0$
is the zero-frequency limit of the scattering rate. The
superscript indicates that this is the first guess for $\Sig$. The
Green's function components are computed, which gives the explicit
form of Eq.~(\ref{eq:sigma}). Upon evaluating the numerics, it is
seen that this first iteration generates a nonzero (real and
negative) term for $\Sigma_{B1}$. So, the diagonal self-energy
assumption turns out to be inconsistent, in contrast to the
situation for $\psi=0$. We then modify our guess, assuming
self-energy of the form
$\Sig^{(2)}=-i\Gamma_0\widetilde{(\sigma_0\otimes\tau_0)}-B_1\widetilde{(\sigma_1\otimes\tau_1)}$.
The Green's function is computed again, using Dyson's equation,
and the self-energy equations are obtained explicitly. It is noted
that the symmetries of Eq.~(\ref{eq:greensymmetries}) still hold.
Again, the equations (\ref{eq:Self}) are solved iteratively; the
result is a non-zero $\Sigma_{B3}$ component as well. Once again,
the Green's functions are modified to incorporate this term, and
the iterative scheme is applied. Calculation of the self-energy
based on the assumption
\bea
 \Sig^{(3)}=-i\Gamma_0\widetilde{(\sigma_0\otimes\tau_0)}-B_1\widetilde{(\sigma_1\otimes\tau_1)}-B_3\widetilde{(\sigma_1\otimes\tau_3)}\nonumber\\
 \Gamma_0,B_1,B_3>0
\eea generates $\Gamma_0$, $B_1$, and $B_3$ that are much larger
than any remaining terms, and hence provides the self-consistent
values of $\Sigma_{A0},\Sigma_{B1}$ and $\Sigma_{B3}$. A plot of
the 6 components of $\Sig$ is displayed in
Fig.~\ref{fig:6sigmasBW} for a representative parameter set, where
we see that the three terms of the ansatz are indeed dominant. For
the remainder of this paper, the effect of the $\Sigma_{A1}$,
$\Sigma_{A3}$ and $\Sigma_{B0}$ components will be ignored. The
self-consistent Green's functions are provided in
Appendix~\ref{app:greensfunctions}, while additional details of
the self-energy calculation are discussed in
Appendix~\ref{app:divergence}.

%%%%%%%%%%%%%%%%%%%%%%%%%%%%%%%%%%%%%%%%%%%%%%%%%%%%%%%%%%%%%%%%%%%%%%%%%%%%%%%%%%%%%%%%%%%%%%%%%%%%%%%%%%%%%%%%%%%%%%
\subsection{SCBA Results}
\label{sec:SCBAresults}
%%%%%%%%%%%%%%%%%%%%%%%%%%%%%%%%%%%%%%%%%%%%%%%%%%%%%%%%%%%%%%%%%%%%%%%%%%%%%%%%%%%%%%%%%%%%%%%%%%%%%%%%%%%%%%%%%%%%%%
In order to discuss the numerical results contained in this paper,
it is necessary to make a note about the units employed. The
following discussion of units applies as well to the numerical
analysis of the results of the thermal conductivity calculation in
Sec.~\ref{sec:analysis}. Because we are studying the evolution of
the system with respect to increasing CDW order parameter $\psi$,
we wish to express energies in units of $\psi_c$, the value of
$\psi$ which gaps the clean system. In order to do this, the
cutoff $p_0$ is fixed such that the Brillouin zone being
integrated over in Eq.~(\ref{eq:intparametrization}) has the
correct area. In this way, $p_0$ sets the scale of the product
$v_f v_\Delta$; a parameter
$\beta\equiv\sqrt{\frac{v_f}{v_\Delta}}$ is defined to represent
the velocity anisotropy. Then,
$\frac{p_0}{\psi_c}=\frac{\pi}{2a}\sqrt{v_fv_\Delta}$, so that we
may eliminate the frequently occurring parameter $4\pi
v_fv_\Delta$ by expressing lengths in units of
$\frac{4}{\sqrt{\pi}}a\approx 2.25 a$. Impurity density
$n_{\mathrm{imp}}$ is thus recast in terms of impurity fractions
$z$ according to $n_{\mathrm{imp}}=\frac{16}{\pi}z$. Finally, the
parameters of the scattering potential are recast in terms of
their anisotropy. We define $V_2\equiv R_2 V_1$ and $V_3\equiv R_3
V_1$.

With these modifications, the original set of parameters,
$\{n_i,V_1,V_2,V_3,v_f,v_\Delta,p_0,\psi,\psi_c\}$ is reduced to
$\{z,V_1,R_2,R_3,\beta,p_0,\psi\}$. For the work contained herein,
the cutoff $p_0$ is fixed at $p_0=100$. The self-energy in the
self-consistent Born approximation was computed for different
scattering potentials as a function of impurity fraction and CDW
order parameter $\psi$. Since it was found that three of the
components, $\Sigma_{A0}$, $\Sigma_{B1}$ and $\Sigma_{B3}$,
dominate over the others, we will subsequently analyze only those
three components, referring to their magnitudes as $\Gamma_0$,
$B_1$, and $B_3$ respectively.

As $z\rightarrow 0$, the Green's functions become impossibly
peaked from a numerical point of view. For sufficiently large $z$,
depending on the strength of the scatterers, the Born
approximation breaks down. Given a scattering strength of
$V_1=110$, cutoff $p_0=100$, scattering potentials that fall off
slowly in $k$-space and velocity anisotropy ratios $\beta\equiv
\sqrt{v_f/v_\Delta}={1,2,3,4}$, this puts the range of $z$ in
which our numerics may be applied at roughly between one half and
one percent.

\begin{figure}
    \resizebox{3.25in}{!}{\includegraphics{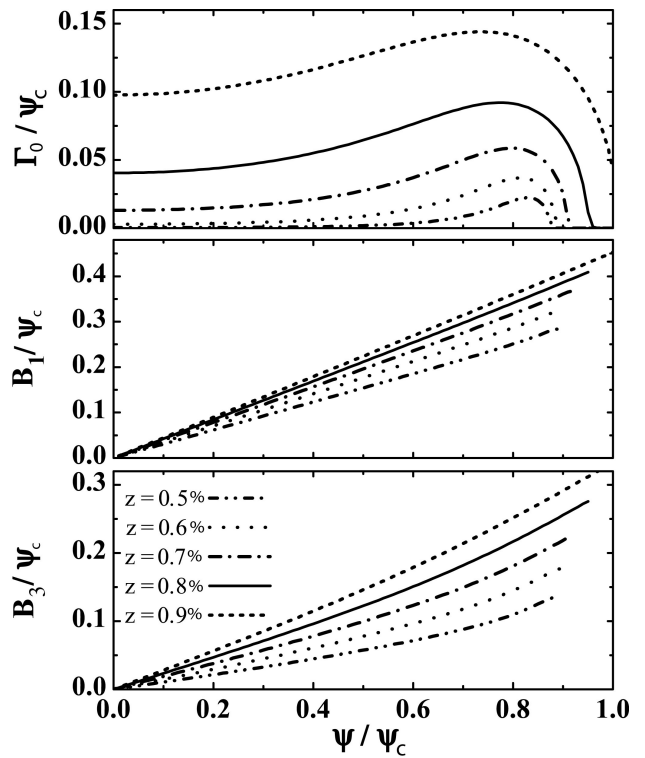}}
  \caption{Effect of disorder on charge-order-dependence of
  self-energy components. To satisfy Dyson's equation, it is necessary
  to include three (extended-Nambu space) components of the self energy. Their
  self-consistent values are plotted here for several different values
  of impurity fraction $z$. Here, the scattering potential is given in
  our three parameter model as $\{V_1,R_2,R_3\}=\{110,0.9,0.8\}$, which
  represents a fairly short-ranged potential. These results are for
  the case of isotropic nodes ($v_f=v_\Delta$). All energies are in units
  of $\psi_c$.}
  \label{fig:sigmaofPsiBeta1}
\end{figure}

\begin{figure}
    \centerline{\resizebox{3.25in}{!}{\includegraphics{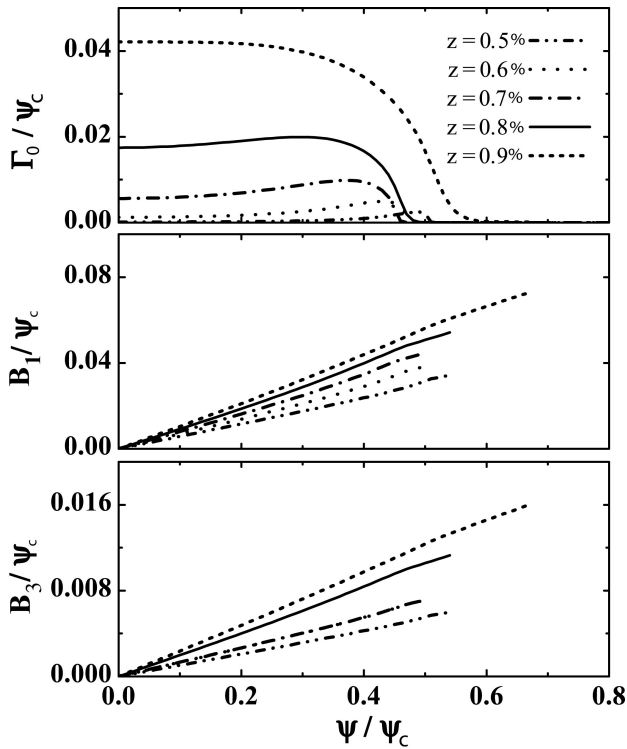}}}
  \caption{Effect of disorder on charge-order-dependence of self-energy
  components. This figure illustrates the case where $v_f=16 v_\Delta$.
  The scattering potential is again given by $\{V_1,R_2,R_3\}=\{110,0.9,0.8\}$,
  representing a fairly short-ranged potential. The plots for $B_1$
  and $B_3$ terminate before $\psi$ reaches $\psi_c$ because for
  sufficiently large $\psi$, the excitations become gapped and our nodal
  approximations break down.}
  \label{fig:sigmaofPsiBeta4}
\end{figure}

\begin{figure}[h]
    \centerline{\resizebox{3 in}{!}{\includegraphics{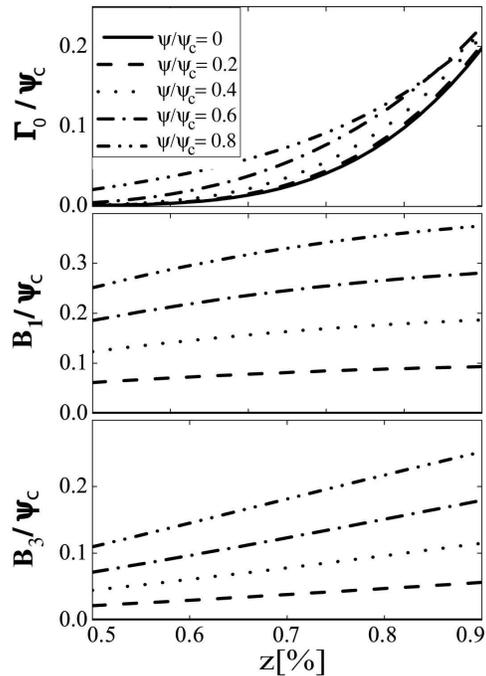}}}
  \caption{Effect of charge order on disorder-dependence of self-energy components.
  Nonzero components of $\wt{\Sigma}(z)$
  are shown for impurity fraction $z$ ranging from $0.5$ to $1.0\%$,
  for charge order parameter $\psi$=0, 0.2, 0.4, 0.6, and 0.8 (in units of $\psi_c$).
  These results are for scattering parameters $\{V_1,R_2,R_3\}=\{110,0.9,0.8\}$
  and $v_f=v_\Delta$. Similar results are obtained for the case of anisotropic
  nodes.}
  \label{fig:sigmaofz}
\end{figure}

Some results for $\wt{\Sigma}(\psi)$, for several values of $z$,
are shown in Figs.~\ref{fig:sigmaofPsiBeta1} and
\ref{fig:sigmaofPsiBeta4}. These plots correspond to the same
parameters, except that Fig.~\ref{fig:sigmaofPsiBeta1} illustrates
the $v_f=v_\Delta$ case, and Fig.~\ref{fig:sigmaofPsiBeta4}
illustrates $v_f=16v_\Delta$. In all cases it is seen that \bea
 B_1(\psi,z)\simeq b_1(z)\psi\nonumber\\
 B_3(\psi,z)\simeq b_3(z)\psi
\eea where the dependence of $B_1$, $B_3$, $b_1$, and $b_3$ on the
remaining parameters is implicit. For much of the parameter space
sampled, $\Gamma_0$ does not have much $\psi$ dependence, except
that it typically rises and then falls to zero at some
sufficiently large $\psi<\psi_c$. This feature will be revisited
in Sec. \ref{sec:analysis}, wherein it is explained that this
vanishing scattering rate coincides with vanishing thermal
conductivity, and corresponds to the point at which the system
becomes effectively gapped and our nodal approximations break
down. The value of $\psi$ at which this occurs depends on the
entire set of parameters used, and will be referred to as
$\psi_c^*$.
The observed $z$ dependence is not very surprising, in light of
Eq.~(\ref{eq:Self}). The self-energy components depend on $z$
roughly according to
\bea
 \Gamma_0 &\sim &p_0 \exp{(-\frac{1}{z})}\nonumber\\
 B_1 &\sim & z\nonumber\\
 B_3&\sim &z
\eea as can be seen in Fig.~\ref{fig:sigmaofz}. When $\psi=0$,
$\Gamma_0$ is given by the closed-form expression obtained in
Ref.~\onlinecite{dur01}, $\Gamma_0=p_0 \exp(\frac{-1}{2\pi c})$,
where $c=\frac{n_i(V_1^2+2V_2^2+V_3^2)}{8\pi^2v_fv_\Delta}$. For
finite $\psi$, this precise form does not hold, but the strong $z$
dependence of $\Gamma_0$ remains, in contrast to that of $B_1$ and
$B_3$. Note that the $z$ dependence of $B_1$ and $B_3$ is roughly
linear for $\psi \ll \psi_c^*$. As $\psi$ approaches $\psi_c^*$
the functions diverge slightly from linearity. Results for several
values of $\psi<\psi_c^*$ are shown in the figure.

%%%%%%%%%%%%%%%%%%%%%%%%%%%%%%%%%%%%%%%%%%%%%%%%%%%%%%%%%%%%%%%%%%%%%%%%%%%%%%%%%%%%%%%%%%%%%%%%%%%%%%%%%%%%%%%%%%%%%%
\section{Thermal Conductivity}
\label{sec:thermalconductivity}
%%%%%%%%%%%%%%%%%%%%%%%%%%%%%%%%%%%%%%%%%%%%%%%%%%%%%%%%%%%%%%%%%%%%%%%%%%%%%%%%%%%%%%%%%%%%%%%%%%%%%%%%%%%%%%%%%%%%%%
Thermal conductivity was calculated using the Kubo formula
\cite{mah01,Fetter},
\be
 \frac{\kappa(\Omega,T)}{T}=-\frac{\mathrm{Im}\Pi_{\mathrm{Ret}}(\Omega)}{\Omega\,\,T^2},
\ee
where $\Pi_{\mathrm{Ret}}(\Omega)$ is the retarded thermal
current-current correlation function. To find this correlator, it
is necessary to first compute the appropriate thermal current
operator. For our model hamiltonian, this is done in Ref.
\onlinecite{dur02} with the result \be
 \label{eq:current}
 \vect{\widetilde{j}}^\kappa_0=\lim_{\substack{q\rightarrow0\\
 \Omega\rightarrow0}}\sum_{\substack{k,\omega}}
 (\omega+\frac{\Omega}{2})\psi^\dagger_k
 \left(\widetilde{\vect{v}}_{fM}+\widetilde{\vect{v}}_{\Delta M}\right)\psi_{k+q},
\ee
where a generalized velocity is defined as
\bea
 \widetilde{\vect{v}}_{\alpha M}=v_\alpha^x \widetilde{M}_\alpha^x\hat{x}
 +v_\alpha^y\widetilde{M}_\alpha^y\hat{y}\nonumber\\
 \widetilde{M}_\alpha^x\equiv\widetilde{(\sigma_3\otimes\tau_\alpha)}
 \hspace{15pt} \widetilde{M}_\alpha^y\equiv\widetilde{(\sigma_0\otimes\tau_\alpha)}
\eea where $\alpha=\{f,\Delta\}$ and
$\tau_\alpha=\{\tau_3,\tau_1\}$ for Fermi and gap velocities
respectively.

\begin{figure}[ht]
   \subfigure[]{
   \centerline{\resizebox{3.25in}{!}{\includegraphics{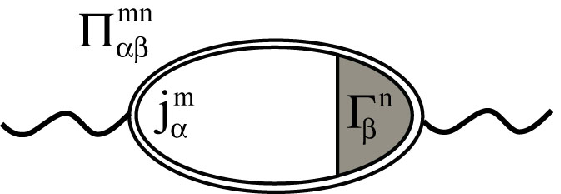}}}
   \label{fig:bubbleFeynman}
   }
    \subfigure[]{
   \centerline{\resizebox{3.25in}{!}{\includegraphics{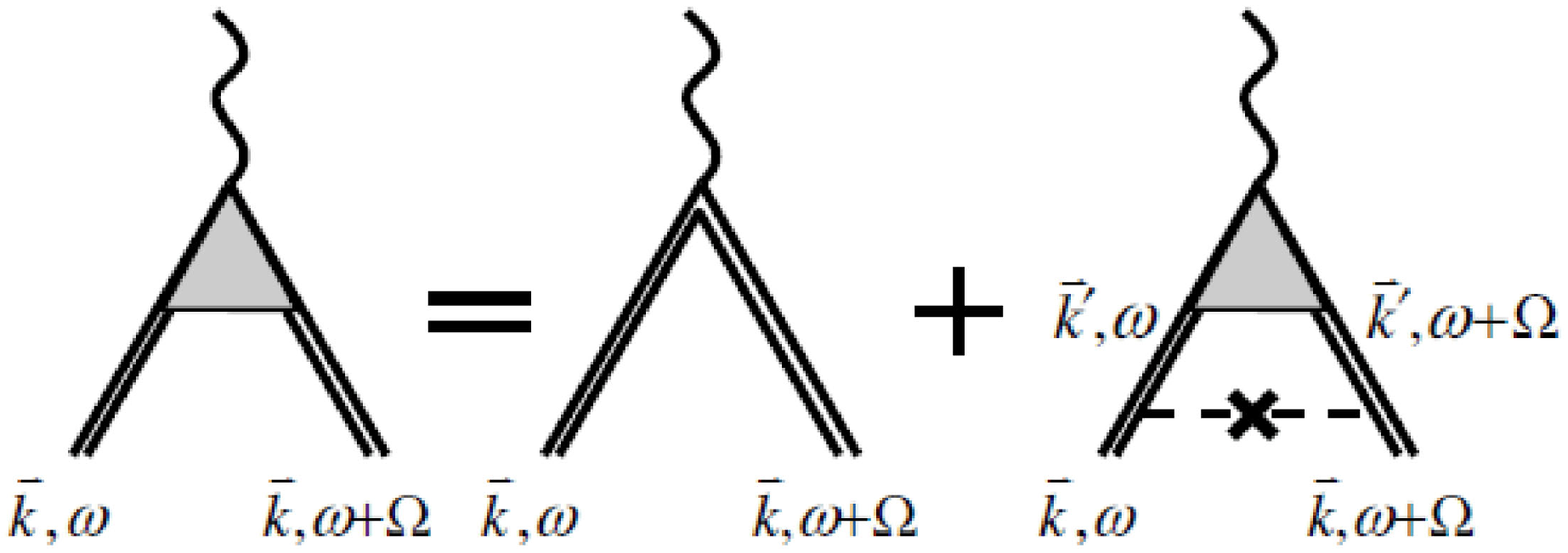}}}
   \label{fig:vertexFeynman}
   }
   \caption{
   (a) Feynman diagram representing the correlation function
   $\Pi_{\alpha\beta}^{\mathrm{mn}}$ in terms of a bare vertex
   $j_\alpha^{\mathrm{m}}$, and a dressed vertex $\Gamma_\beta^{\mathrm{n}}$.
   (b) Feynman diagram representing the (ladder series) dressed vertex in
   terms of the bare vertex and the Born scattering event.
   }
\end{figure}

To calculate a thermal conductivity that satisfies Ward
identities, vertex corrections must be included on the same
footing as the self-energy corrections to the single particle
Green's function. The details of this calculation are similar to
those performed in Appendix B of Ref.~\onlinecite{dur01}. The
impurity scattering diagrams which contribute to the ladder series
of diagrams are included by expressing the correlation function in
terms of a dressed vertex, as shown in
Fig.~\ref{fig:bubbleFeynman}.
The current-current correlation function is obtained from this
dressed bubble. The bare current operator of
Eq.~(\ref{eq:current}) is associated with one vertex of the
bubble, while the dressed vertex of Fig.~\ref{fig:vertexFeynman}
is associated with the other. Evaluating
Fig.~\ref{fig:bubbleFeynman}, we find that the current-current
correlation function takes the form \bea
 \Pi^{mn}(i\Omega)=\sum_{\alpha,\beta=f,\Delta}\Pi^{mn}_{\alpha\beta}(i\Omega)\nonumber\\
 \Pi^{mn}_{\alpha\beta}(i\Omega)=\frac{1}{k_B T}\sum_{i \omega}(i\omega+\frac{i\Omega}{2})^2 \sum_k\nonumber\\ \mathrm{Tr} \left[\widetilde{\mathcal{G}_1}v_\alpha k_\alpha^m\widetilde{M_\alpha^m}\widetilde{\mathcal{G}_2}v_\beta\widetilde{M_\beta^n}\widetilde{\Gamma_\beta^n}\right]
\eea where
$\widetilde{\mathcal{G}}_1\equiv\widetilde{\mathcal{G}(\vect{k},i\omega})$,
$\widetilde{\mathcal{G}_2}\equiv\widetilde{\mathcal{G}}(\vect{k},i\omega+i\Omega)$,
and
$\widetilde{\Gamma_\beta^n}=\widetilde{\Gamma_\beta^n}(\vect{k},i\omega,i\Omega)$
represents the dressed vertex depicted in
Fig.~\ref{fig:vertexFeynman}. The Greek indices denote ``Fermi''
and ``gap'' terms, while the Roman indices denote the position
space components of the tensor. We use
Fig.~\ref{fig:vertexFeynman} to find the form of the vertex
equation, and then make the ansatz that \be
 \widetilde{\vect{\Gamma}_\beta}(\vect{k},i\omega,i\Omega)=\Big(\wt{\openone}+\wt{\Lambda}(|\vect{k}|,i\omega,i\Omega)\Big)\hat{k},
\ee
which leads to the scalar equations
\bea
 \widetilde{\Gamma}_\beta^n(\vect{k},i\omega,i\Omega)=k_n(\widetilde{\openone}+\widetilde{\Lambda}_\beta^n).
\eea
Looking for solutions of this form, we see that the scalar vertex function is
\be
 \wt{\Lambda}_\beta^n=\mathrm{n}_{i}\sum_{k'}\wt{M}_\beta^n\wt{V}_{kk'}\wt{\mathcal{G}}_2\wt{M}_\beta^n(\wt{\openone}+\wt{\Lambda}_\beta^n)
 \wt{\mathcal{G}}_1\wt{V}_{k'k}\frac{k^{'n}_\beta}{k^n_\beta}.
\ee Since we are working with nodal quasiparticles, we utilize the
parametrization of Eq.~(\ref{eq:intparametrization}), so that the
vertex function is now a function of node index $j$ and local
momentum $\vect{p}$ \bea
 \wt{\Lambda}_\beta^n&=&n_{\mathrm{imp}}\sum_{j'=1}^4\underline{V}_{jj'}\underline{V}_{j'j}
 (\frac{k_{\beta n}^{(j')}}{k_{\beta n}^{(j)}})\int\frac{\mathrm{d}^2p'}{8\pi^2 v_fv_2}\nonumber\\& &\wt{M}_\beta^n(\wt{\sigma_0\otimes\tau_3})
 \wt{\mathcal{G}}_2\wt{M}_\beta^n(\wt{\openone}+\wt{\Lambda}_\beta^n)\wt{\mathcal{G}}_1(\wt{\sigma_0\otimes\tau_3}).
\eea
Arbitrarily choosing $j=1$, then for $j'=\{1,2,3,4\}$
\bea
 \frac{k_{1x}^{(j')}}{k_{1x}^{(1)}}=\{1,-1,-1,1\} \hspace{20pt} \frac{k_{1y}^{(j')}}{k_{1y}^{(1)}}=\{1,1,-1,-1\} \nonumber\\
 \frac{k_{2x}^{(j')}}{k_{2x}^{(1)}}=\{1,-1,-1,1\} \hspace{20pt} \frac{k_{2y}^{(j')}}{k_{2y}^{(1)}}=\{1,1,-1,-1\}.
\eea Using the node space matrix representing the 3-parameter
scattering potential \be
 \underline{V}_{jj'}=
  \begin{pmatrix}
  V_1 & V_2 & V_3 & V_2 \\
  V_2 & V_1 & V_2 & V_3 \\
  V_3 & V_2 & V_1 & V_2 \\
  V_2 & V_3 & V_2 & V_1
  \end{pmatrix}
\ee
we obtain for the vertex equation
\be
 \label{eq:vertex}
 \wt{\Lambda}^n_\beta=\gamma \int \frac{\mathrm{d}^2p'}{\pi}\wt{M}_\beta^n(\wt{\sigma_0\otimes\tau_3})\wt{\mathcal{G}}_2
 \wt{M}_\beta^n(\wt{\openone}+\wt{\Lambda}_\beta^n)\wt{\mathcal{G}}_1(\wt{\sigma_0\otimes\tau_3})
\ee where $\gamma\equiv n_{\mathrm{imp}}\frac{V_1^2-V_3^2}{8\pi
v_fv_2}$. The correlator then becomes
\bea
 \Pi_{\alpha \beta}^{mn}(i\Omega)&=&
 v_\alpha v_\beta \frac{1}{\beta}\sum_{i\omega}(i\omega+\frac{i\Omega}{2})^2\sum_k (k_{\alpha m}k_{\beta n})
 \nonumber\\ & &\mathrm{Tr}\left(\wt{\mathcal{G}}_1 \wt{M}_\alpha^m\wt{\mathcal{G}}_2\wt{M}_\beta ^n(\wt{\openone}+\wt{\Lambda}_\beta ^n)\right)
 \nonumber\\
 &=&v_\alpha v_\beta \frac{1}{\beta}\sum_{i\omega}(i\omega+\frac{i\Omega}{2})^2 \sum_{j=1}^4(k_{\alpha m}^{(j)}k_{\beta n}^{(j)})
 \nonumber\\ & &\int \frac{\mathrm{d}^2p}{8\pi^2v_fv_\Delta}
 \mathrm{Tr}\left(\wt{\mathcal{G}}_1 \wt{M}_\alpha^m\wt{\mathcal{G}}_2\wt{M}_\beta ^n(\wt{\openone}+\wt{\Lambda}_\beta ^n)\right).
\eea
Since
\be
 \sum_{j=1}^4k_{\alpha m}^{(j)}k_{\beta n}^{(j)}=2\left( (1-\delta_{\alpha \beta})\eta_m+\delta_{\alpha \beta}\right)\delta_{mn}
\ee
we can write
\be
 \label{eq:correlator}
 \Pi_{\alpha\beta}^{mn}(i\Omega)=2\pi c_{\alpha\beta}^{mn}\frac{1}{\beta}\sum_{i\omega}(i\omega+\frac{i\Omega}{2})^2
 \mathrm{Tr}\left(\wt{I}_{\alpha\beta}^{mn}(\wt{\openone}+\wt{\Lambda}_\beta^n)\right)
\ee
where
\bea
 \label{eq:cmn}
 c_{\alpha \beta}^{mn}&\equiv&\frac{1}{8\pi^2}\frac{v_\alpha v_\beta}{v_f v_\Delta}\Big((1-\delta_{\alpha\beta})\eta_m+
 \delta_{\alpha\beta}\Big)\delta_{mn}
\eea
and
\bea
 \label{eq:integral}
 \wt{I}_{\alpha\beta}^{mn}(i\omega,i\omega+i\Omega)&\equiv&\int\frac{\mathrm{d}^2p}{\pi}\wt{\mathcal{G}}_1\wt{M}_\alpha^m\wt{\mathcal{G}}_2\wt{M}_\beta^n.
\eea To calculate the conductivity, we will need
Tr$(\wt{I}_{\alpha\beta}^{m})$ and
Tr$(\wt{I}_{\alpha\beta}^m\wt{\Lambda}_\beta^n)$. For $\psi=0$, it
is possible to compute the integral in Eq.~(\ref{eq:integral})
analytically, but for general $\psi$ we had to compute the
integrals numerically. We note that if we write \be
 \wt{I}=
 \begin{pmatrix}
  I_A & I_B \\
  I_C & I_D
 \end{pmatrix},
\ee apply the symmetry properties of
Eq.~(\ref{eq:greensymmetries}) and reverse the order of
integration of $p_1$ and $p_2$, then $I_A=I_D$, and $I_B=I_C$, so
that the most general expansion of $\wt{I}_{\alpha\beta}^{mn}$ in
Nambu space is \be
 \wt{I}_{\alpha\beta}^{mn}=\sum_{i=0}^1\sum_{i'=0}^3(I_{\alpha\beta}^{mn})_{ii'}(\wt{\sigma_i}\otimes\tau_{i'}).
\ee
Then
\bea
 \mathrm{Tr}(\wt{I}_{\alpha\beta}^{mn})&=&\mathrm{Tr}\left(
 \sum_{i=0}^1\sum_{i'=0}^3(I_{\alpha\beta}^{mn})_{ii'}(\wt{\sigma_i\otimes\tau_{i'}})\right)\nonumber\\
 &=& 4(I_{\alpha\beta}^{mn})_{00},
\eea
while if we use the same expansion for
\bea
 \wt{\Lambda}_\beta^n=\sum_{i=0}^1\sum_{i'=0}^3(\Lambda_\beta^n)_{ii'}(\wt{\sigma_i
 \otimes\tau_{i'}}),
\eea
we find
\bea
 \mathrm{Tr}(\wt{I}_{\alpha\beta}^{mn}\wt{\Lambda}_\beta^n)=\sum_{ij=0}^1\sum_{i'j'=0}^3(I_{\alpha\beta}^{mn})_{ii'}
 (\Lambda_\beta^n)_{jj'}\nonumber\\
 \mathrm{Tr}(\wt{\sigma_i\sigma_j\otimes\tau_{i'}\tau_{j'}})\nonumber\\
 =4\sum_{i=0}^1\sum_{i'=0}^3(I_{\alpha\beta}^{mn})_{ii'}(\Lambda_\beta^n)_{ii'}.
\eea
Then Eq.~(\ref{eq:vertex}) becomes
\bea
 4(\Lambda_\beta^n)_{ii'}=\mathrm{Tr}\left((\wt{\sigma_i\otimes\tau_{i'}})\wt{\Lambda}_\beta^n\right)\nonumber\\
 =\gamma\int\frac{\mathrm{d}^2p}{\pi}\mathrm{Tr}((\wt{\sigma_i\otimes\tau_{i'}})\wt{M}_\beta^n(\wt{\sigma_0\otimes\tau_3})
 \\ \wt{\mathcal{G}}_2\wt{M}_\beta^n(\wt{\openone}+\wt{\Lambda_\beta^n})\wt{\mathcal{G}}_1(\wt{\sigma_0\otimes\tau_3}))\nonumber\\
 = \gamma\mathrm{Tr}\left(\wt{L}_{\beta ii'}^n(\wt{1}+\wt{\Lambda}_\beta^n)\right)
\eea
where
\bea
 \label{eq:integralprime}
 \wt{L}_{\beta ii'}^n\equiv\int\frac{\mathrm{d}^2p}{\pi}\wt{\mathcal{G}}_1(\wt{\sigma_0\otimes\tau_3})(\wt{\sigma_i\otimes\tau_{i'}})\nonumber\\\wt{M}_\beta^n
 (\wt{\sigma_0\otimes\tau_3})\wt{\mathcal{G}}_2\wt{M}_\beta^n.
\eea
The symmetries of $\wt{\mathcal{G}}$ which were used to see
which components of $\wt{I}_{\alpha\beta}^{mn}$ were $0$ can also
be applied to $\wt{L}_{\beta ii'}^n$ with the result that
$(L_{\beta ii'}^n)_A=(L_{\beta ii'}^n)_D$, $(L_{\beta
ii'}^n)_B=\eta_i (L_{\beta ii'}^n)_C$, where
$\eta_i=\left\{\begin{array}{c} + 1, \,i=0,1\\-1,\, i=2,3
\end{array}\right\}$. Since all that is required for the
conductivity is $i=0,1$, we use the expansion
\be
 \wt{L}_{\beta ii'}^n=\sum_{j=0}^1\sum_{j'=0}^3(\wt{\sigma_j\otimes\tau_{j'}})(L_{\beta ii'}^n)_{jj'}
\ee
so that
\bea
 \label{eq:vertexfinal}
 (\Lambda_\beta^n)_{ii'}=\frac{1}{4}\gamma\mathrm{Tr}\left(\wt{L}_{\beta ii'}^n(\wt{\openone}+\wt{\Lambda}_\beta^n)\right)\nonumber\\
 =\frac{1}{4}\gamma\,\,\mathrm{Tr}\,(\sum_{j=0}^1\sum_{j'=0}^3(L_{\beta ii'}^n)_{jj'}(\wt{\sigma_j\otimes\tau_{j'}})\nonumber\\
 +\sum_{jk=0}^1\sum_{j'k'=0}^3(L_{\beta ii'})_{jj'}(\Lambda_\beta^n)_{kk'})\nonumber\\
 =\gamma\left((L_{\beta ii'}^n)_{00}+\sum_{j=0}^1\sum_{j'=0}^3(L_{\beta ii'}^n)_{jj'}(\Lambda_\beta^n)_{jj'}\right).
\eea
The thermal conductivity is obtained from the retarded
current-current correlation function
\bea
 \frac{\kappa^{mn}(\Omega)}{T}=-\frac{1}{T}\frac{\mathrm{Im}\left(\Pi_{\mathrm{ret}}^{mn}(\Omega)\right)}{\Omega},
\eea where
$\Pi_{\mathrm{ret}}(\Omega)=\Pi(i\Omega\rightarrow\Omega+i\delta).$
To get the retarded correlator we first perform the Matsubara
summation. Consider the summand of Eq.~\ref{eq:correlator}, which
we redefine according to
\be
 \label{eq:jeqn}
 J(i\omega,i\omega+i\Omega)=\mathrm{Tr}\left(\wt{I}_{\alpha\beta}^{mn}(\wt{\openone}+\wt{\Lambda}_\beta^n)\right).
\ee The function $J(i\omega,i\omega+i\Omega)$ is of the form
$J(i\omega,i\omega+i\Omega)=f(A(i\omega)B(i\omega+i\Omega))$ where
$A$ and $B$ are dressed Green's functions of a complex variable
$z=i\omega_n$, so that $J$ is analytic with branch cuts occurring
where $z$ and $z+i\Omega$ are real. The Matsubara summation needed
is performed by integrating on a circular path of infinite radius,
so that the only contribution is from just above and just below
the branch cuts, \bea
 \Pi_{\alpha\beta}^{mn}=-c_{\alpha\beta}^{mn}\frac{1}{i}\oint \mathrm{d}z (z+\frac{i\Omega}{2})^2J(z,z+i\Omega)\nonumber\\
 =-c_{\alpha\beta}^{mn}\frac{1}{i}\int_{-\infty}^\infty\mathrm{d}\epsilon\, n_f(\epsilon)\Big( \nonumber\\
 (\epsilon+\frac{i\Omega}{2})^2
 (J(\epsilon+i\delta,\epsilon+i\Omega)-J(\epsilon-i\delta,\epsilon+i\Omega))\nonumber\\+(\epsilon-\frac{i\Omega}{2})^2(
 J(\epsilon-i\Omega,\epsilon+i\delta)-J(\epsilon-i\Omega,\epsilon-i\delta))\Big).
\eea To obtain the retarded function, we analytically continue
$i\Omega\rightarrow\Omega+i\delta$. Then we let
$\epsilon\rightarrow\epsilon+\Omega$ in the third and fourth
terms, so that \bea
 \Pi_{\alpha\beta}^{mn}(\Omega)_{\mathrm{ret}}&=&c_{\alpha\beta}^{mn}\int_{-\infty}^\infty\mathrm{d}\epsilon\,
 n_f(\epsilon+\Omega)-n_f(\epsilon))(\epsilon+\frac{\Omega}{2})^2\nonumber\\
 & &\times \mathrm{Re}\Big(J_{\alpha\beta}^{AR}(\epsilon,\epsilon+\Omega)-J_{\alpha\beta}^{RR}(\epsilon,\epsilon+\Omega)\Big)
\eea where $J^{AR}$ and $J^{RR}$ are defined by Eqs.
(\ref{eq:jeqn}) and (\ref{eq:vertexfinal}) and are composed of the
universal-limit Green's functions given in Appendix
\ref{app:greensfunctions}. Taking the imaginary part, we find
\bea
 \frac{\kappa^{mn}(\Omega,T)}{T}&=&-\int_{-\infty}^\infty \mathrm{d}\epsilon\frac{n_f(\epsilon+\Omega)-n_f(\epsilon)}{\Omega}
 \left(\frac{\epsilon+\frac{\Omega}{2}}{T}\right)^2\nonumber\\
 \sum_{\alpha\beta}c_{\alpha\beta}^{mn}&\mathrm{Re}&(J_{\alpha\beta}^{AR}(\epsilon,\epsilon+\Omega))-J_{\alpha\beta}^{RR}(\epsilon,\epsilon+\Omega)).
\eea
In taking the $\Omega\rightarrow 0$ limit, the difference in
Fermi functions becomes a derivative. Evaluating the integral,
$\int\mathrm{d}\epsilon(-\frac{\mathrm{d}n}{\mathrm{d}\epsilon})(\frac{\epsilon}
{T})^2=\frac{\pi^2k_B^2}{3}$, we find that
\bea
 \frac{\kappa_{\alpha\beta}^{mm}(0,0)}{T}=\frac{\pi^2k_B^2}{3}c_{\alpha\beta}^{mm}\mathrm{Re}\left(J_{\alpha\beta}^{AR}(0,0)-J_{\alpha\beta}^{RR}
 (0,0)\right).
\eea
That $\kappa^{xy}=\kappa^{yx}=0$ is seen from
Eq.~(\ref{eq:cmn}). Finally, since the $\alpha\neq\beta$ integrals
are traceless, the result for the thermal conductivity is
\bea
 \label{eq:thermal}
 \frac{\kappa^{mm}}{T}=\frac{k_B^2}{3}\frac{v_f^2+v_\Delta^2}{v_fv_\Delta}\frac{1}{8}\left(J_{\alpha\beta}^{AR}(0,0)-J_{\alpha\beta}^{RR}
 (0,0)\right).
\eea

%%%%%%%%%%%%%%%%%%%%%%%%%%%%%%%%%%%%%%%%%%%%%%%%%%%%%%%%%%%%%%%%%%%%%%%%%%%%%%%%%%%%%%%%%%%%%%%%%%%%%%%%%%%%%%%%%%%%%%
\section{Results}
\label{sec:analysis}
%%%%%%%%%%%%%%%%%%%%%%%%%%%%%%%%%%%%%%%%%%%%%%%%%%%%%%%%%%%%%%%%%%%%%%%%%%%%%%%%%%%%%%%%%%%%%%%%%%%%%%%%%%%%%%%%%%%%%%
For a discussion of the units employed in the analysis, one can
refer to Sec.~\ref{sec:SCBAresults}. The reduced set of parameters
for the model is $\{z,V_1,R_2,R_3,\beta,p_0,\psi\}$. We explored a
limited region of this parameter space, calculating the integrals
and solving the matrix equation numerically. In particular, we
looked at the $\psi$ dependence of $\kappa$. To vary the
anisotropy of the scattering potential, we considered the
$\{R_2,R_3\}$ values of $\{0.9,0.8\}$, $\{0.7,0.6\}$, and
$\{0.5,0.3\}$, and kept fixed the constant $c$ (given after
Eq.~(\ref{eq:Self}))by appropriately modifying $V_1$. For
$\{R_2,R_3\}=\{0.9,0.8\}$, we used $V_1=110$. The rationale for
keeping $c$ fixed is that the self-energy depends only on $c$,
$\beta$ and $p_0$. Additionally, we explored the dependence of the
thermal conductivity on impurity fraction $z$ and velocity
anisotropy $\beta$. For all computations we set the cutoff
$p_0=100$; this simply fixes a particular value of the product
$v_f v_\Delta$ for these calculations.

%%%%%%%%%%%%%%%%%%%%%%%%%%%%%%%%%%%%%%%%%%%%%%%%%%%%%%%%%%%%%%%%%%%%%%%%%%%%%%%%%%%%%%%%%%%%%%%%%%%%%%%%%%%%%%%%%%%%%%
\subsection{Vertex Corrections}
%%%%%%%%%%%%%%%%%%%%%%%%%%%%%%%%%%%%%%%%%%%%%%%%%%%%%%%%%%%%%%%%%%%%%%%%%%%%%%%%%%%%%%%%%%%%%%%%%%%%%%%%%%%%%%%%%%%%%%
\begin{figure}
 \centerline{\resizebox{3.25in}{!}{\includegraphics{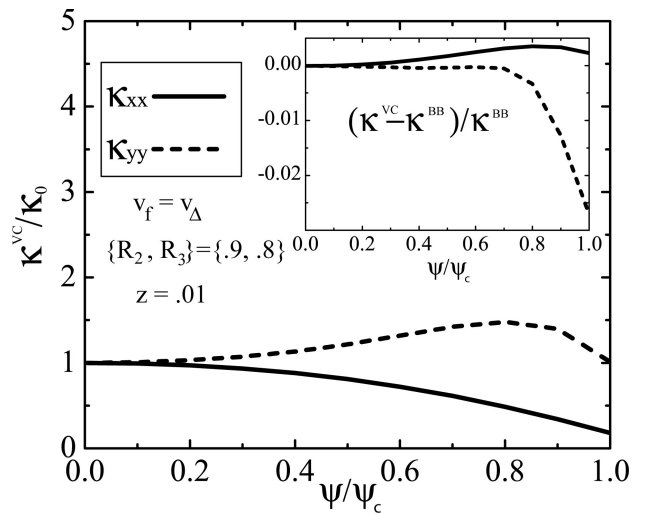}}}
  \caption{Vertex-corrected thermal conductivity, in units of the universal
  conductivity $\kappa_0/T \equiv \frac{k_B}{3\hbar}(v_f/v_\Delta+v_\Delta/v_f)$.
  This data reflects a short range scattering potential
$\{V_1,R_2,R_3\}=\{110,0.9,0.8\}$, impurity fraction $z$=0.01, and
isotropic Dirac quasiparticles ($v_f=v_\Delta$). The inset
displays the discrepancy between the bare-bubble and
vertex-corrected results, in units of the bare-bubble result. It
is clear that the vertex corrections are of little quantitative
importance for these particular parameters.
   }
  \label{fig:vertex78}
\end{figure}
\begin{figure}
    \centerline{\resizebox{3.25in}{!}{\includegraphics{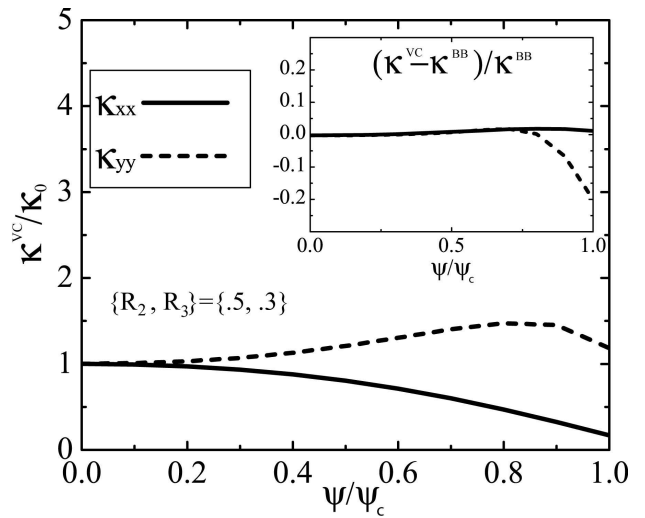}}}
 \caption{Vertex-corrected thermal conductivity, in units of the universal
 conductivity $\kappa_0/T \equiv \frac{k_B}{3\hbar}(v_f/v_\Delta+v_\Delta/v_f)$.
 This figure portrays
the effect that a different scattering potential has on the
importance of vertex corrections. Here, a longer range potential
$\{V_1,R_2,R_3\}=\{140,0.5,0.3\}$ was used, again with impurity
fraction $z$=$0.01$ and $v_f=v_\Delta$. The inset displays the
discrepancy between the bare-bubble and vertex-corrected results,
in units of the bare-bubble result. From this, we determine that
vertex corrections make a more substantial correction as the
forward scattering limit is approached, but only once the charge
ordering is quite strong.
   }
  \label{fig:vertex60}
\end{figure}
\begin{figure}
 \centerline{\resizebox{3.25in}{!}{\includegraphics{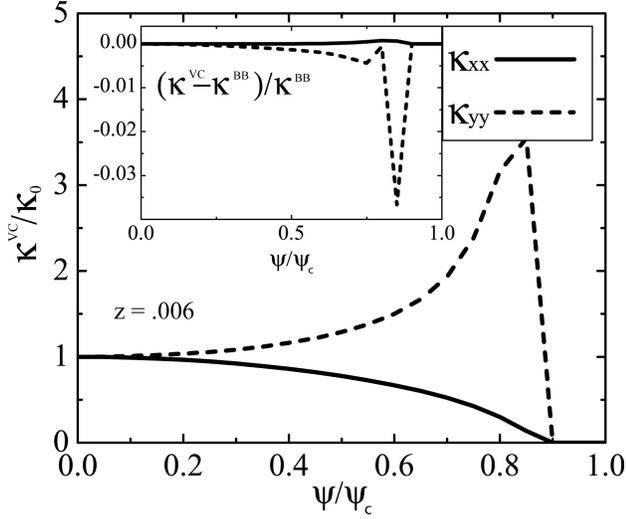}}}
  \caption{Vertex-corrected thermal conductivity, in units
  of the universal conductivity $\kappa_0/T \equiv \frac{k_B}{3\hbar}(v_f/v_\Delta+v_\Delta/v_f)$.
  Again, a short-ranged
  scattering potential, $\{V_1,R_2,R_3\}=\{110,0.9,0.8\}$ and isotropic
  nodes ($v_f=v_\Delta$) are used.  This figure displays the effect of a
  smaller impurity fraction than that depicted in Fig.\ref{fig:vertex78}.
  The inset displays the discrepancy between the bare-bubble and vertex-corrected
  results, in units of the bare-bubble result; since the scattering potential
  falls off slowly (in k-space) here, the vertex corrections are again
  quite unimportant.}
  \label{fig:vertex48}
\end{figure}
\begin{figure}
    \centerline{\resizebox{3.25in}{!}{\includegraphics{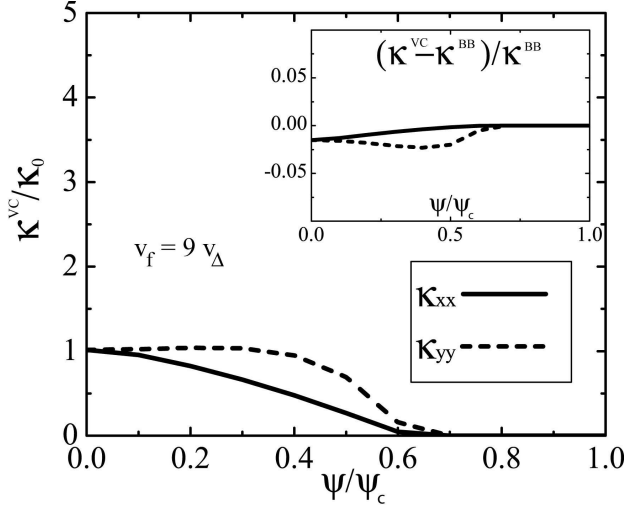}}}
  \caption{Vertex-corrected thermal conductivity, in units of the universal
  conductivity $\kappa_0/T \equiv \frac{k_B}{3\hbar}(v_f/v_\Delta+v_\Delta/v_f)$,
  for short-ranged scattering
  potential, $\{V_1,R_2,R_3\}=\{110,0.9,0.8\}$ and impurity fraction $z=0.01$.
  These calculations differ from those of Fig.\ref{fig:vertex78} in that they
  apply to the case of a more anisotropic Dirac spectrum with $v_f=9v_\Delta$. The thermal
  conductivity has a qualitatively similar $\psi$ dependence, but vanishes for
  a smaller value of $\psi$ than for the isotropic case. The inset displays
  the discrepancy between the bare-bubble and vertex-corrected results, in
  units of the bare-bubble result; again, the vertex corrections do not
  significantly modify the bare-bubble results.}
  \label{fig:vertex58}
\end{figure}
The importance of including the vertex corrections is determined
by comparing the vertex corrected thermal conductivity with that
of the bare-bubble. If
$\frac{\kappa^{VC}-\kappa^{BB}}{\kappa^{BB}}<<1$ for a region of
parameter space, then in that regime the bare-bubble results can
be used instead. This is of threefold practicality: the
bare-bubble results are less computationally expensive, the
bare-bubble expression is much simpler to analyze, and other
hamiltonians could be more easily studied.

The bare bubble thermal conductivity can be obtained by setting
$\wt{\Lambda}_\beta^n\rightarrow \wt{0}$ in Eq.~(\ref{eq:jeqn}),
or by using a spectral representation, as in
Ref~\onlinecite{dur02}; both methods have the same result. For
impurity fraction $z$ ranging from 0.5$\%$ to $1\%$, the
importance of the vertex corrections is largely seen to be
negligible, which implies that an analysis of the bare bubble
results is sufficient.

Figs. \ref{fig:vertex78}-\ref{fig:vertex58} illustrate the vertex
corrected thermal conductivities, $\kappa^{VC}$, in the main
graphs, while the insets display the relative discrepancy with
respect to the bare bubble thermal conductivities
$\frac{\kappa^{VC}-\kappa^{BB}}{\kappa^{BB}}$. Each is plotted as
a function of the amplitude of the CDW, $\psi/\psi_c$, where
$\psi_c$ indicates the maximal CDW for which the clean system
remains gapless. We will postpone analysis of the character of the
thermal conductivity until Sec.~V~C.

To gauge the importance of the vertex corrections, we look first
at Fig.~\ref{fig:vertex78}. The inset indicates that the vertex
corrections do not signifigantly modify the bare bubble thermal
conductivity. Although their importance grows somewhat with
increasing $\psi$, the correction is still slight.

Next, Fig.~\ref{fig:vertex78} is used as a reference against which
to consider the dependence of vertex corrections on scattering
potential, impurity fraction, and velocity anisotropy. The next
three figures are the results of computations with each of these
parameters modified in turn. By comparing Fig.~\ref{fig:vertex60}
with Fig.~\ref{fig:vertex78} we conclude that the vertex
corrections become more important when the scattering potential is
peaked in $k$-space, but are unimportant for potentials that fall
off slowly in $k$-space.

Fig.~\ref{fig:vertex78} and Fig.~\ref{fig:vertex48} correspond
roughly to the largest and smallest $z$ for which these
calculations are valid. Comparison of these two figures, as well
as that of intermediary values of $z$ (not displayed) indicates
that the relative importance of the vertex corrections is
independent of $z$. Nor does increasing the velocity anisotropy
affect their importance, as seen by making a comparison between
Fig.~\ref{fig:vertex78} and Fig.~\ref{fig:vertex58}.

%%%%%%%%%%%%%%%%%%%%%%%%%%%%%%%%%%%%%%%%%%%%%%%%%%%%%%%%%%%%%%%%%%%%%%%%%%%%%%%%%%%%%%%%%%%%%%%%%%%%%%%%%%%%%%%%%%%%%%
\subsection{Clean Limit Analysis}
%%%%%%%%%%%%%%%%%%%%%%%%%%%%%%%%%%%%%%%%%%%%%%%%%%%%%%%%%%%%%%%%%%%%%%%%%%%%%%%%%%%%%%%%%%%%%%%%%%%%%%%%%%%%%%%%%%%%%%
It is of great interest to consider the behavior of the thermal
conductivity in the clean $(z\rightarrow 0)$ limit. Because the
thermal conductivity is composed of integrals over
$\vect{p}$-space of functions which become increasingly peaked in
this limit, there exists a sufficiently small $z$ beyond which it
is not possible to perform the requisite numerical integrations.
However, it is still possible to obtain information about this
regime. To that end, we will examine the form of the bare-bubble
thermal conductivity, and consider the $z\rightarrow 0$ limit. As
we shall see, this will enable us to determine the value of $\psi$
at which the nodal approximation, and hence this calculation, is
no longer valid. Additionally, a closed-form result for the
thermal conductivity in the $z\rightarrow 0$ limit is obtained for
the isotropic ($v_f=v_\Delta$) case. The bare-bubble thermal
conductivity, identical with setting $\wt{\Lambda}\rightarrow
\wt{0}$ in Eq.~(\ref{eq:thermal}), is
\begin{widetext}
 \bea
  \label{eq:cleanThermal}
  \kappa^{mm}&=&\frac{k_B}{3}\frac{v_f^2+v_2^2}{v_f v_2}J^m \hspace{50pt} J^m=\int\frac{\mathrm{d}^2\vect{p}}{2\pi}\frac{N_1+N_2}{D}\hspace{50pt}\epsilon_1\equiv\epsilon_k\hspace{50pt}\Delta_1\equiv\Delta_k\nonumber\\
  N_1&=&A\left((A+B+\epsilon_1^2+\Delta_1^2)^2+(A+B+\epsilon_2^2+\Delta_2^2)^2\right)
  \hspace{41pt}\epsilon_2\equiv\epsilon_{k+G}\hspace{38pt}\Delta_2\equiv\Delta_{k+G}\nonumber\\
  N_2&=&\eta_m A\Big((\psi-B_3)^2((\epsilon_1+\epsilon_2)^2-(\Delta_1-\Delta_2)^2)+B_1^2((\Delta_1+\Delta_2)^2-(\epsilon_1-\epsilon_2)^2)
  -4B_1(\psi-B_3)(\epsilon_1\Delta_1+\epsilon_2\Delta_2))\Big)\nonumber\\
  D&=&\Big[ (A+B+\epsilon_1^2+\Delta_1^2)(A+B+\epsilon_2^2+\Delta_2^2)-B\Big((\epsilon_1+\epsilon_2)^2+(\Delta_1-\Delta_2)^2\Big)\nonumber\\
  & &+4 B_1\Big(B_1(\epsilon_1\epsilon_2-\Delta_1\Delta_2)+(\psi-B_3)(\epsilon_1\Delta_2+\epsilon_2\Delta_1)\Big)\Big]^2,
 \eea
\end{widetext}
where $A\equiv\Gamma_0^2$ and $B\equiv(\psi-B_3)^2+B_1^2$. Since
the results of Section~\ref{sec:SCBAresults} indicated that
$\Gamma_0\sim \exp{(-\frac{1}{z})}$ and $B_1,B_3\sim z$, in the
$z\rightarrow 0$ limit, $A\rightarrow 0$ much faster than
$B_1\rightarrow 0$ or $B_3\rightarrow 0$. Therefore in taking the
$z\rightarrow 0$ limit we will first let $A\rightarrow 0$ to
obtain a result still expressed in terms of $B_1$ and $B_3$. The
denominator can be rearranged as
\begin{widetext}
 \bea
  D&=&\Big( (A^2+A(2B+\epsilon_1^2+\Delta_1^2+\epsilon_2^2+\Delta_2^2)+f\Big)^2 \hspace{20pt} \mathrm{where}\nonumber\\
  f&=&B^2+(\epsilon_1^2+\Delta_1^2)(\epsilon_2^2+\Delta_2^2)-2B(\epsilon_1\epsilon_2-\Delta_1\Delta_2)+4\Big(B_1(\epsilon_1\epsilon_2-\Delta_1\Delta_2)
  +(\psi-B_3)(\epsilon_1\Delta_2+\epsilon_2\Delta_1)\Big)\nonumber\\
  &=&\Big((\epsilon_1 \epsilon_2-\Delta_1\Delta_2) - (2B_1^2-B) \Big)^2 +\Big( (\epsilon_1\Delta_2+\epsilon_2\Delta_1) + 2B_1(\psi-B_3)\Big)^2
 \eea
\end{widetext}
We are thus considering, in the limit that $A\rightarrow0$, an
integral of the form
\be
 \int\mathrm{d}^2\vect{p}\frac{A\,g(\vect{p})}{\big(A h(\vect{p})+f(\vect{p})\big)^2}
\ee Note that any nonzero contribution to this integral must come
from a region in $\vect{p}$-space in which $f(\vect{p})=0$. We
will consider separately the isotropic case ($v_f=v_\Delta$) and
the anisotropic case ($v_f > v_\Delta$).

%%%%%%%%%%%%%%%%%%%%%%%%%%%%%%%%%%%%%%%%%%%%%%%%%%%%%%%%%%%%%%%%%%%%%%%%%%%%%%%%%%%%%%%%%%%%%%%%%%%%%%%%%%%%%%%%%%%%%%
\subsubsection{Isotropic Case}
%%%%%%%%%%%%%%%%%%%%%%%%%%%%%%%%%%%%%%%%%%%%%%%%%%%%%%%%%%%%%%%%%%%%%%%%%%%%%%%%%%%%%%%%%%%%%%%%%%%%%%%%%%%%%%%%%%%%%%
For the special case where $v_f=v_\Delta$, it is possible to
calculate the integral of Eq.~(\ref{eq:cleanThermal}) exactly, by
taking the $A\rightarrow0$ limit, and choosing another
parametrization. The coordinates
$q_1\equiv\epsilon_k-\epsilon_{k+Q}$ and
$q_2\equiv\epsilon_k+\epsilon_{k+Q}-1$, have their origin located
at the midpoint of the white and gray dots of
Fig.~\ref{fig:BZfig}. Using these coordinates, in the
$A\rightarrow0$ limit we find that the elements of
Eq.~(\ref{eq:cleanThermal}) become
\begin{widetext}
 \bea
  \label{eq:cleanprogress}
  N_1&=&2A\Big(B^2+B(q^2+1)+\frac{1}{4}(q^2+1)^2+q^2-q_2^2\Big)\nonumber\\
  N_2&=&2\eta_mA\Big((\psi-B_3)^2((\epsilon_1+\epsilon_2)^2-(\Delta_1-\Delta_2)^2)+B_1^2((\Delta_1+\Delta_2)^2-(\epsilon_1-\epsilon_2)^2)-4B_1
  (\psi-B_3)(\epsilon_1\Delta_2+\epsilon_2\Delta_1)\Big)\nonumber\\
  &=&2\eta_m\Big((\psi-B_3)^2(q_2^2+2q_2+1-q_1^2)+B_1^2(q_2^2-2q_2+1-q_1^2)-4B_1(\psi-B_3)(2q_2^2-q^2-1)\Big)\nonumber\\
  &=&2\eta_m A\Big[(2q_2^2-q^2+1)\Big((\psi-B_3)^2-2B_1(\psi-B_3)+B_1^2\Big)+2q_2\Big((\psi-B_3)^2-B_1^2\Big)+4B_1(\psi-B_3)\Big]\nonumber\\
  D&=&\Big[ 2A\Big(1+B-2B_1(\psi-B_3)\Big)+\Big(q_2-(\psi^2-B_1^2)^2\Big)^2+\frac{1}{4}\Big(q^2-(1-4B_1(\psi-B_3))\Big)^2\Big]^2.
 \eea
\end{widetext}
Now the part of the denominator not proportional to $A$, the
$f$-term, is zero when
\be
 \label{eq:conditions}
 q_2=(\psi-B_3)^2-B_1^2 \hspace{10pt} \mathrm{and} \hspace{10pt} q^2=1-4B_1(\psi-B_3).
\ee
\begin{figure}[h]
    \centerline{\resizebox{3.25in}{!}{\includegraphics{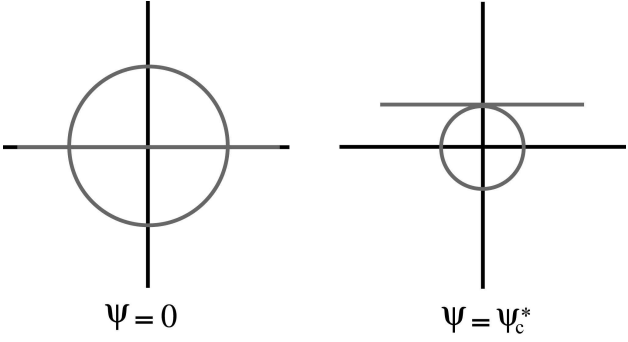}}}
  \caption{Illustrated is a schematic view of the line and circle whose
  intersection determines whether gapless excitations remain, for the
  isotropic case ($v_f=v_\Delta$). The left figure indicates the situation
  in the absence of charge ordering, that is, for $\psi=0$, where the radius
  of the circle is 1 and the line lies on the horizontal axis. The right figure
  indicates the situation at $\psi=\psi_c^*$,
  when charge ordering is such that the excitation spectrum becomes gapped.
  In the clean case, the $\psi$ evolution corresponds to moving the line past
  the circle. With self-consistent disorder, the radius of the circle and height
  of the line are both functions of $\psi$; in each instance, this construction
  can be used to determine the value of $\psi$ at which the quasiparticle spectrum
  becomes gapped. This value of $\psi$ is referred to as $\psi_c^*$ in this paper.}
  \label{fig:lineCircle}
\end{figure}
In $q_1/q_2$ coordinates, these are the equations of a horizontal
line and a circle, which must intersect for there to be a nonzero
contribution to the integral, since each term is positive
definite. In the simplified disorder treatment of
Ref.~\onlinecite{dur02} for which $B_1=B_3=0$ and
$\Gamma_0=\mathrm{constant}$, these constraints simplify to
$q_2=\psi^2$ and $q^2=1$, so that no contribution occurs when
$\psi>1$ (Note that as in the numerical analysis, $\psi$, being an
energy, is measured in units of $\psi_c$). With the
self-consistent treatment of disorder, there will likewise be a
sufficiently large value of $\psi$ beyond which the line and
circle no longer intersect; we will call this value $\psi_c^*$
(see Fig.~\ref{fig:lineCircle}). We interpret $\psi_c^*$ as the
point beyond which the system becomes effectively gapped. This is
consistent with the exact result found by computing the
eigenvalues of the completely clean hamiltonian (as
$\psi_c^*=\psi_c$ in that case).

In Sec.~\ref{sec:SCBAresults} it was determined that $B_1\simeq
b_1 \psi$ and $B_3\simeq b_3 \psi$, where $b_1$ and $b_3$ depend
on the remaining parameters of the model. Using this approximate
form for $B_1$ and $B_3$, the condition for the maximum $\psi$ for
which the constraints of Eq.~(\ref{eq:conditions}) are satisfied,
\be
 1-4B_1(\psi-B_3)=\Big((\psi-B_3)^2-B_1^2\Big)^2,
\ee
indicates that
\be
 \psi_c^{*2}\simeq \frac{\pm\Big((1-b_3)\mp b_1\Big)^2}{\Big((1-b_3-b_1)(1-b_3+b_1)\Big)^2}.
\ee
Since $\psi_c^{*2}>0$, we find that for $v_f=v_\Delta$,
\be
 \label{eq:psicrit}
 \psi_c^* \simeq \frac{1}{1-b_3+b_1}.
\ee We now proceed with the calculation of the clean-limit thermal
conductivity. Substituting the conditions of
Eq.~(\ref{eq:conditions}) into Eq.~(\ref{eq:cleanprogress}), we
find that the numerators become
\begin{widetext}
 \bea
  N_1 &=& 4A\Big[\Big(1-2B_1(\psi-B_3)\Big)\Big(1+B-2B_1(\psi-B_3)\Big)\Big]\nonumber\\
  N_2 &=& 4\eta_m A\Big(1+B-2B_1(\psi-B_3)\Big)\Big([(\psi-B_3)^2-B_1^2]^2+2B_1(\psi-B_3)\Big)
 \eea
\end{widetext}
both of which are independent of $\vect{q}$, so that the clean
limit result hinges upon the integral
\bea
 I=\int\frac{\mathrm{d}^2q}{4\pi}\frac{A}{\Big(k_1 A+(q_2-k_2)^2+\frac{1}{4}(q^2-k_3)^2\Big)^2},
\eea
where
\bea
 k_1&=&2\Big(1+B-2B_1(\psi-B_3)\Big)\nonumber\\
 k_2&=&(\psi-B_3)^2-B_1^2\nonumber\\
 k_3&=&1-4B_1(\psi-B_3).
\eea
The details of this integration are reported in Appendix
\ref{app:integration}, with the result \be
 I=\frac{1}{2k_1\sqrt{k_3-k_1^2}}.
\ee
We can now write the anisotropic clean limit thermal conductivity
\begin{widetext}
 \bea
  J&=&\frac{1-2B_1\psi+\eta_m\Big([(\psi-B_3)^2-B_1^2]^2+2B_1(\psi-B_3)\Big)}{\sqrt{1-4B_1(\psi-B_3)-[(\psi-B_3)^2-B_1^2]^2}}
  \hspace{3pt}\Theta\Big(1-4B_1(\psi-B_3)-[(\psi-B_3)^2-B_1^2]^2\Big)\nonumber\\
  J^{xx}&=&\sqrt{1-4B_1(\psi-B_3)-[(\psi-B_3)^2-B_1^2]^2}\hspace{6pt} \Theta \Big(1-4B_1(\psi-B_3)-[(\psi-B_3)^2-B_1^2]^2\Big)\nonumber\\
  J^{yy}&=&\frac{1+[(\psi-B_3)^2-B_1^2]^2}{\sqrt{1-4B_1(\psi-B_3)-[(\psi-B_3)^2-B_1^2]^2}}\hspace{6pt}\Theta\Big(1-4B_1(\psi-B_3)-[(\psi-B_3)^2-B_1^2]^2\Big),
 \eea
\end{widetext}
where the $\Theta$ function is the Heaviside step function. Using
the definition for $\psi_c^*$ found in Eq.~(\ref{eq:psicrit}), and
defining
\be
 \label{eq:chicrit}
 \chi \equiv \frac{1}{1-b_3-b_1},
\ee
we are able to rewrite the dimensionless conductivity in
terms
of parameters easily extrapolated from SCBA calculations
\begin{widetext}
 \bea
  \label{eq:cleanlimitthermal}
  J^{xx}&=&\frac{\kappa^{xx}}{\kappa_0}=\sqrt{\Big( 1-\frac{\psi^2}{\psi_c^{*2}} \Big) \Big( 1+\frac{\psi^2}{\chi^2} \Big)} \hspace{5pt}
  \Theta[\Big(1-\frac{\psi^2}{\psi_c^{*2}}\Big)]\nonumber\\
  J^{yy}&=&\frac{\kappa^{yy}}{\kappa_0}=\Big(1+\frac{\psi^4}{\psi_c^{*2}\chi^2}\Big)\Big(1-\frac{\psi^2}{\psi_c^{*2}}\Big)^{-1/2}\Big(1+\frac{\psi^2}{\chi^2}\Big)^{-1/2}
  \hspace{6pt}\Theta[\Big(1-\frac{\psi^2}{\psi_c^{*2}}\Big)]
 \eea
\end{widetext}
in which form it is clear that the thermal conductivity vanishes
for $\psi>\psi_c^*$.

%%%%%%%%%%%%%%%%%%%%%%%%%%%%%%%%%%%%%%%%%%%%%%%%%%%%%%%%%%%%%%%%%%%%%%%%%%%%%%%%%%%%%%%%%%%%%%%%%%%%%%%%%%%%%%%%%%%%%%
\subsubsection{Anisotropic Case}
\begin{figure}
    \centerline{\resizebox{3.25in}{!}{\includegraphics{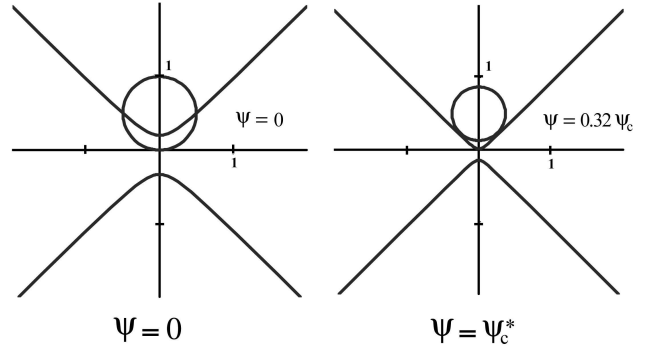}}}
  \caption{For generally anisotropic Dirac quasiparticles,
  the construction used in Fig.\ref{fig:lineCircle} is modified to contain
  a hyperbola and circle. When these no longer intersect, the excitation
  spectrum becomes gapped. Illustrated is the construction for scattering
  parameter values $\{V_1,R_2,R_3\}=\{110,0.9,0.8\}$, impurity fraction
  $z=0.01$, and with $v_f=4v_\Delta$. For these parameters it was determined
  that the value of $\psi$ at which the spectrum becomes gapped is given
  by $\psi_c^*=0.32\psi_c$.}
  \label{fig:hyperbolaCircle}
\end{figure}

%%%%%%%%%%%%%%%%%%%%%%%%%%%%%%%%%%%%%%%%%%%%%%%%%%%%%%%%%%%%%%%%%%%%%%%%%%%%%%%%%%%%%%%%%%%%%%%%%%%%%%%%%%%%%%%%%%%%%%
For the case of anisotropic nodes, $v_f > v_\Delta$, the integral
of Eq.~(\ref{eq:cleanThermal}) becomes intractable. However, it is
still possible to predict $\psi_c^*$. Using the same $q_1/q_2$
coordinates, the $f$-part of the denominator is again a sum of two
positive definite terms. Again, the only contributions to the
clean-limit thermal conductivity arise when $f=0$, which again
provides two equations \bea
 \label{eq:2eqns}
 x^2+(y-a)^2=R^2\nonumber\\
 (y-b)^2-x^2=c^2
\eea
where
\bea
 a&=&\frac{1}{\beta}(\beta-1)\nonumber\\
 b&=&\frac{\beta^4-2\beta^3-1}{\beta^4-1}\nonumber\\
 c&=&\frac{2\beta}{\beta^4-1}\sqrt{1-(\beta^4-1)\Big((\psi-B_3)^2-B_1^2\Big)}\nonumber\\
 R&=&\sqrt{(1-\frac{1}{\beta}(\beta-1))^2-4B_1(\psi-B_3)}.
\eea
This defines a hyperbola and a circle, again parametrized by
$\psi$. One instance of this is depicted in
Fig.~\ref{fig:hyperbolaCircle}.
The value of $\psi$ at which these equations no longer have a
solution is $\psi_c^*$. The computed values for $\psi_c^*$ are
included for comparison in the plots of thermal conductivity in
Fig.~\ref{fig:kappa1} and Fig.~\ref{fig:kappa4}.

%%%%%%%%%%%%%%%%%%%%%%%%%%%%%%%%%%%%%%%%%%%%%%%%%%%%%%%%%%%%%%%%%%%%%%%%%%%%%%%%%%%%%%%%%%%%%%%%%%%%%%%%%%%%%%%%%%%%%%
\subsection{Effect of Self-Consistent Disorder}
%%%%%%%%%%%%%%%%%%%%%%%%%%%%%%%%%%%%%%%%%%%%%%%%%%%%%%%%%%%%%%%%%%%%%%%%%%%%%%%%%%%%%%%%%%%%%%%%%%%%%%%%%%%%%%%%%%%%%%
\begin{figure}
 \centerline{\resizebox{3.25 in}{!}{\includegraphics{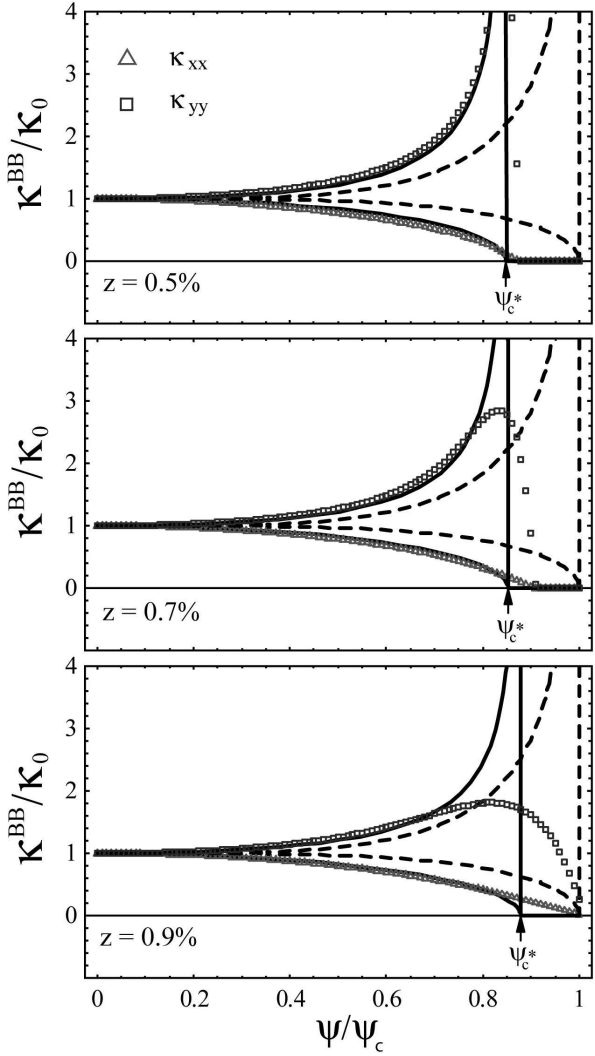}}}
 \caption{Effects of disorder on the charge-order-dependence of the bare-bubble thermal
 conductivity, isotropic case ($v_f=v_\Delta$).
 Note how an increase in the impurity fraction, $z$, broadens out the peak in the
 conductivity. As the disorder becomes sufficiently small, the computed
 conductivity (triangles and squares) attains a limiting value that closely
 agrees with the closed-form clean-limit results of Eq.~(\ref{eq:cleanlimitthermal})
 (shown with solid lines).
 The thermal conductivity obtained by simply letting $\wt{\Sigma}\rightarrow$-i$\Gamma_0$
 (as in Ref.~\onlinecite{dur02}) is shown with dashed lines. The effect of the
 self-consistent disorder is to renormalize the effective $\psi$ at which the thermal
 conductivity vanishes (from $\psi_c$ to $\psi_c^*$).  Here, we have considered short-ranged
 scatterers $\{V_1,R_2,R_3\}=\{110,0.9,0.8\}$.}
 \label{fig:kappa1}
\end{figure}
\begin{figure}
 \centerline{\resizebox{3.25 in}{!}{\includegraphics{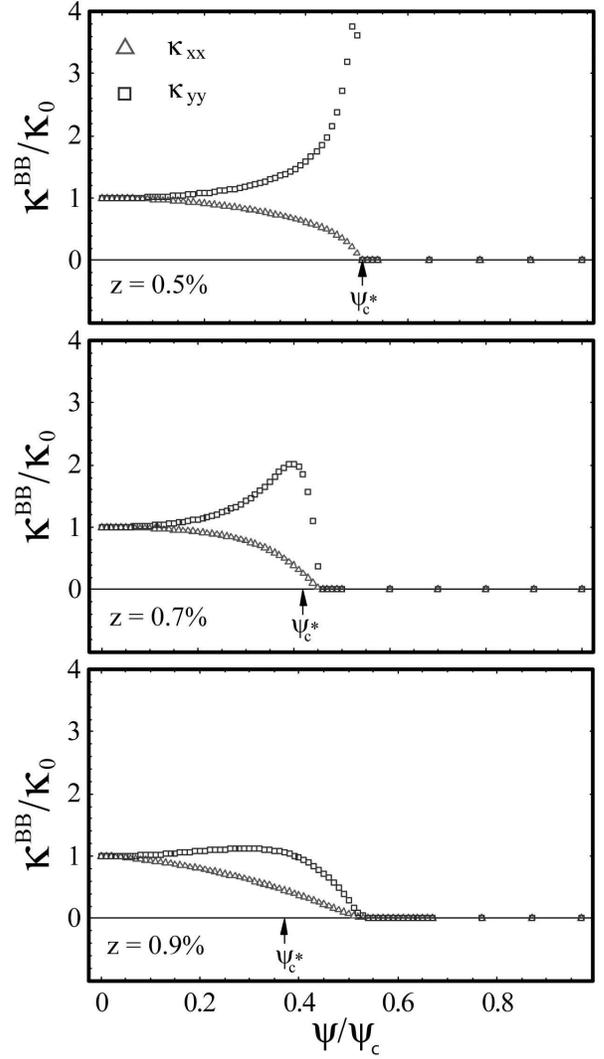}}}
  \caption{Effects of disorder on the charge-order-dependence of the bare-bubble
  thermal conductivity, anisotropic case ($v_f=16v_\Delta$).  The effect of disorder is the
same as in the isotropic case, which is to mix gapped and gapless
states, smearing the peak in $\kappa_{yy}$ across the renormalized
nodal transition point, $\psi_c^*$. It is interesting to note that
for this anisotropic case, $\psi_c^*$ is significantly smaller
than $\psi_c$. Again, we have considered short-ranged scatterers
$\{V_1,R_2,R_3\}=\{110,0.9,0.8\}$.}
 \label{fig:kappa4}
\end{figure}
Satisfied that vertex corrections are of little importance, we set
about analyzing the form of the thermal conductivity by studying
the bare-bubble results. Thermal conductivity $\kappa$ was
computed for $\beta\equiv\sqrt{v_f/v_\Delta}$ values of $1,2,3$
and $4$ (that is, for $v_f/v_\Delta$=$1,4,9$ and $16$). In
Fig.~\ref{fig:kappa1} is presented a representative plot of
$\kappa$ for $v_f=v_\Delta$. The clean limit prediction for
$\kappa$ (Eq.~(\ref{eq:cleanlimitthermal})) is computed by fitting
$b_1$ and $b_3$ from the self-energy calculations. These clean
limit predictions are then plotted on the same graph with the
numerical results of the thermal conductivity for the same
parameters.  In addition, the clean limit results of the simpler
disorder model of Ref.~\onlinecite{dur02} are also shown for the
$v_f=v_\Delta$ case. Increasing disorder broadens the peak in
$\kappa^{yy}$ near $\psi_c^*$. For $z=0.005$, the numerical
computation is already almost exactly given by the clean limit
results, while for $z=0.009$, the features of the conductivity are
nearly totally smeared out, as seen in Fig.~\ref{fig:kappa1}. In
this figure, the value of $\psi_c^*$ given by
Eq.~(\ref{eq:psicrit}) is indicated with an arrow.

For $v_f>v_\Delta$, the thermal conductivity has the same
characteristics as for $v_f=v_\Delta$, except that $\psi_c^*$ is
generally smaller for larger $\beta$. The numerically computed
thermal conductivities for the case of $\beta=4$ are shown in
Fig.~\ref{fig:kappa4}. In this figure, the value of $\psi_c^*$ is
computed by determining the largest value of $\psi$ for which
Eqs.~(\ref{eq:2eqns}) have a solution, and is indicated with an
arrow. It is clear from these graphs that the self-consistent
disorder renormalizes the amplitude of charge density wave at
which the thermal conductivity vanishes, and that the amount of
renormalization is heavily dependent on the velocity anisotropy
ratio, and varies only slightly with changing impurity fraction.

%%%%%%%%%%%%%%%%%%%%%%%%%%%%%%%%%%%%%%%%%%%%%%%%%%%%%%%%%%%%%%%%%%%%%%%%%%%%%%%%%%%%%%%%%%%%%%%%%%%%%%%%%%%%%%%%%%%%%%%%%%%%%
\section{Conclusions}
\label{sec:conc}
%%%%%%%%%%%%%%%%%%%%%%%%%%%%%%%%%%%%%%%%%%%%%%%%%%%%%%%%%%%%%%%%%%%%%%%%%%%%%%%%%%%%%%%%%%%%%%%%%%%%%%%%%%%%%%%%%%%%%%%%%%%%%
The work described in this paper investigates the low temperature
thermal conductivity of a $d$-wave superconductor with coexisting
charge order in the presence of impurity scattering. We improve
upon the model studied in Ref.~\onlinecite{dur02} by incorporating
the effect of vertex corrections, and by including disorder in a
self-consistent manner. Inclusion of vertex corrections does not
significantly modify the bare-bubble results for short range
scattering potentials. The role vertex corrections play increases
somewhat for longer range scattering potentials, in particular as
the amplitude of charge ordering increases. Nonetheless, for
reasonable parameter values, the inclusion of vertex corrections
is not found to significantly modify the bare-bubble results. This
opens up the possibility of doing bare-bubble calculations for
models with different types of ordering.

Our analysis determined that for self-consistency, it is necessary
to include off-diagonal (in extended-Nambu space) terms in the
self-energy. As the charge ordering increases, the off-diagonal
components become more important, and are found to dominate the
self-energy in the clean limit. We also find that the
zero-temperature thermal conductivity is no longer universal, as
it depends on both disorder and charge order, rather than being
solely determined by the anisotropy of the nodal energy spectrum.

In addition, inclusion of disorder within the self-consistent Born
approximation renormalizes, generally to smaller values, the
critical value of charge ordering strength $\psi$ at which the
system becomes becomes effectively gapped. This renormalization is
seen in the calculated thermal conductivity curves, and depends
primarily on the impurity fraction $z$ and velocity anisotropy
$v_f/v_\Delta$. For larger $v_f/v_\Delta$, the renormalization can
be significant, which may indicate that the calculated effects
could be seen in low-temperature thermal transport even in systems
with relatively weak charge order.

\begin{acknowledgements}
We are grateful to Subir Sachdev for very helpful discussions.
This work is supported by NSF Grant No. DMR-0605919.
\end{acknowledgements}

%%%%%%%%%%%%%%%%%%%%%%%%%%%%%%%%%%%%%%%%%%%%%%%%%%%%%%%%%%%%%%%%%%%%%%%%%%%%%%%%%%%%%%%%%%%%%%%%%%%%%%%%%%%%%%%%%%%%%%%%%%%%%%
\appendix
%%%%%%%%%%%%%%%%%%%%%%%%%%%%%%%%%%%%%%%%%%%%%%%%%%%%%%%%%%%%%%%%%%%%%%%%%%%%%%%%%%%%%%%%%%%%%%%%%%%%%%%%%%%%%%%%%%%%%%%%%%%%%%
\section{Self-consistent Green's functions}
\label{app:greensfunctions}
Here are the Green's functions that
fulfill the self-consistent Born approximation. The superscript
$^{(3)}$ refers to the fact that 3 successive applications of our
self-energy scheme were necessary for self-consistency, as is
explained in Section III.
\begin{widetext}
 \bea
  G_{\mathrm{den}}^{(3)}(\omega)=\left(-f_1^2+f_2^2+f_3^2+(\psi_c+\beta p_1)^2+(\frac{1}{\beta}p_2)^2\right)
  \left(-f_1^2+f_2^2+f_3^2+(\psi_c+\beta p_2)^2+(\frac{1}{\beta}p_1)^2\right)\nonumber\\
  +4\left(f_2^2((\psi_c+\beta p_1)(\psi_c+\beta p_2)-\frac{1}{\beta^2}p_1p_2))-f_3\frac{1}{\beta}((\psi_c+\beta p_1)p_1
  +(\psi_c+\beta p_2)p_2)\right)\nonumber\\
  -(f_2^2+f_3^2)\Big((2\psi_c+\beta(p_1+p_2))^2+\frac{1}{\beta^2}(p_2-p_1)^2\Big)\nonumber
 \eea
 \bea
  \label{eq:SCBAgreenfunctions}
  \GG_{A0}^{(3)}(\omega;p_1,p_2)&=&-f_1\left(-f_1^2+(\psi_c+\beta p_2)^2+\frac{1}{\beta^2}p_1^2+f_2^2+f_3^2\right)\nonumber\\
  \GG_{A1}^{(3)}(\omega;p_1,p_2)&=-&\frac{1}{\beta}p_2\left(-f_1^2+(\psi_c+\beta p_2)^2+(\frac{1}{\beta}p_1)^2\right)-\frac{1}{\beta}p_1(f_3^2-f_2^2)+2(\psi_c+\beta p_2)f_2f_3\nonumber\\
  \GG_{A3}^{(3)}(\omega;p_1,p_2)&=&-(\psi_c+\beta p_1)\left(-f_1^2+(\psi_c+\beta p_2)^2+(\frac{1}{\beta}p_1)^2\right)+(\psi_c+\beta p_2)(f_3^2-f_2^2)+\frac{2}{\beta}p_1f_2f_3\nonumber\\
  \GG_{B0}^{(3)}(\omega;p_1,p_2)&=&f_1\left(f_3(2\psi_c+\beta(p_1+p_2))+f_2\frac{1}{\beta}(p_1+p_2)\right)\nonumber\\
  \GG_{B1}^{(3)}(\omega;p_1,p_2)&=&f_2\left(f_1^2-(\psi_c+\beta p_1)(\psi_c+\beta p_2)+\frac{1}{\beta^2}p_1p_2-f_2^2-f_3^2\right)+f_3\Big((\psi_c+\beta p_1)p_1
  +(\psi_c+\beta p_2)p_2\Big)\nonumber\\
  \GG_{B2}^{(3)}(\omega;p_1,p_2)&=&f_1\left(f_3(\frac{1}{\beta}p_2-\frac{1}{\beta}p_1)+f_2\beta(p_2-p_1)\right)\nonumber\\
  \GG_{B3}^{(3)}(\omega;p_1,p_2)&=&f_3\left(f_1^2-f_2^2-f_3^2+(\psi_c+\beta p_1)(\psi_c+\beta p_2)-\frac{1}{\beta^2}p_1p_2\right)+f_2\Big((\psi_c+\beta p_1)p_1+(\psi_c+\beta p_2)p_2)\Big)\nonumber\\
  \GG_{C0}^{(3)}(\omega;p_1,p_2)&=&\GG_{B0}^{(3)}(\omega;p_1,p_2)\nonumber\\
  \GG_{C1}^{(3)}(\omega;p_1,p_2)&=&\GG_{B1}^{(3)}(\omega;p_1,p_2)\nonumber\\
  \GG_{C2}^{(3)}(\omega;p_1,p_2)&=&-\GG_{B2}^{(3)}(\omega;p_1,p_2)\nonumber\\
  \GG_{C3}^{(3)}(\omega;p_1,p_2)&=&\GG_{B3}^{(3)}(\omega;p_1,p_2)\nonumber\\
  \GG_{D0}^{(3)}(\omega;p_1,p_2)&=&\GG_{A0}^{(3)}(\omega;p_2,p_1)\nonumber\\
  \GG_{D1}^{(3)}(\omega;p_1,p_2)&=&\GG_{C1}^{(3)}(\omega;p_2,p_1)\nonumber\\
  \GG_{D3}^{(3)}(\omega;p_1,p_2)&=&\GG_{C3}^{(3)}(\omega;p_2,p_1)
 \eea
\end{widetext}
To obtain the retarded Green's function $G^{\mathrm{Ret}}(\omega)$
from the above we set \bea
 f_1=\omega-\Sigma_{A0}^{\mathrm{Ret}}(\omega)\hspace{10pt}\nonumber\\
 f_2=\Sigma_{B1}^{\mathrm{Ret}}(\omega)\hspace{10pt}\nonumber\\
 f_3=\psi+\Sigma_{B3}^{\mathrm{Ret}}(\omega).
\eea For the retarded Green's function
$G^{\mathrm{Ret}}(\omega+\Omega)$, we set
$\omega\rightarrow\omega+\Omega$, and for the advanced Green's
function $G^{\mathrm{Adv}}(\omega)$ we set
$\Sigma^{\mathrm{Ret}}\rightarrow\Sigma^{\mathrm{Adv}}$ by taking
the complex conjugate.

%%%%%%%%%%%%%%%%%%%%%%%%%%%%%%%%%%%%%%%%%%%%%%%%%%%%%%%%%%%%%%%%%%%%%%%%%%%%%%%%%%%%%%%%%%%%%%%%%%%%%%%%%%%%%%%%%%%%%%%%%%%%%%
\section{Cutoff-Dependence of Self-Energy}
\label{app:divergence} Here we note that the self-consistent Born
approximation, when applied to the nodal Green's functions used in
this paper, produces a self-energy that contains a logarithmic
divergence, and therefore has a prefactor that is proportional to
the momentum cutoff, set by the size of the Brillouin zone.  By
contrast, the thermal conductivity has no such dependence, and is
therefore a truly nodal property. One difficulty this introduces
is that the prefactor of the self-energy is sensitive to our
choice of coordinates. As the location of the nodes evolves with
charge density wave order parameter $\psi$, computations are
necessarily performed in a different local coordinate system (than
one centered about a node itself). This coordinate shift in the
$p_1$ direction introduces a constant $\Sig_{A3}$ term, even in
the $\psi=0$ instance (whereas using node-centered coordinates,
the anti-symmetric integral is found to vanish). In the $\psi=0$
case, a shift of $\epsilon$ corresponds to the integral \bea
I=\int_{-p_0+\epsilon}^{p_0+\epsilon}\mathrm{d}p_1\int_{-p_0}^{p_0}\mathrm{d}p_2\frac{p_1}{p_1^2+p_2^2+\Gamma_0^2}.
\eea The result is $\pi \epsilon$, which matches the discrepancy.
We therefore subtract off the $\psi=0$ value of $\Sig_{A3}$; the
results shown in Fig.~\ref{fig:6sigmasBW} reflect this
recalibration, as do the subsequent iterations of the self-energy
calculation.

%%%%%%%%%%%%%%%%%%%%%%%%%%%%%%%%%%%%%%%%%%%%%%%%%%%%%%%%%%%%%%%%%%%%%%%%%%%%%%%%%%%%%%%%%%%%%%%%%%%%%%%%%%%%%%%%%%%%%%%%%%%%%%
\section{Calculation of Clean Limit Integral}
\label{app:integration}
For the clean limit of the thermal conductivity we need the integral
\be
\label{eq:integralapp}
 I=\int\frac{\mathrm{d}^2q}{4\pi}\frac{A}{\Big(k_1A+(q_2-k_2)^2+\frac{1}{4}(q^2-k_3)^2\Big)^2},
\ee
in the limit $A\rightarrow 0$. With the substitution
\be
 x_1\equiv x \cos \theta =q_1-1 \hspace{30pt} x_2\equiv x\sin\theta=q_2
\ee
the quantity $Y_1\equiv (q_2-k_2)^2+\frac{1}{4}(q^2-k_3^2)^2$ becomes
\bea
 \label{eq:denominator}
 Y_1&=&\frac{x^4}{4}+k_2^2+\Big(\frac{1-k_3}{2}\Big)^2+\frac{1-k_3}{2}x^2+x^2\nonumber\\& &+(x^2+1-k_3)x \cos\theta-2x k_2\sin\theta.
\eea
To simplify the angular integrand, we get rid of the
$\sin\theta$ term by shifting $\theta\rightarrow\theta +\alpha$.
Then, the last two terms of Eq.~\ref{eq:denominator} become
\begin{widetext}
 \bea
  \frac{x^2+1-k_3}{2}\cos(\theta+\alpha)-k_2\sin(\theta+\alpha)=
  (\frac{x^2+1-k_3}{2}\cos\alpha&-&k_2\sin\alpha)\cos\theta \nonumber\\
 -(\frac{x^2+1-k_3}{2}\sin\alpha+k_2\cos\alpha)\sin\theta.
  \label{eq:terms}
 \eea
 We set the coefficient of the second term on the RHS of Eq.~(\ref{eq:terms}) to $0$, so that the first term becomes
 \bea
 \label{eq:term1}
  -\frac{1}{k_2}\Big( (\frac{x^2+1-k_3}{2})^2+k_2^2\Big)\sin\alpha\cos\theta=
  \frac{-1}{r}\Big(\frac{x^4}{4}+\frac{1-k_3}{2}x^2+(\frac{1-k_3}{2})^2+k_2^2\Big)\cos\theta,
 \eea
 where the RHS of Eq.~(\ref{eq:term1}) is obtained by
 setting $\sin\alpha \equiv k_2/r$, where $r=r(x)$ is an undetermined function of $x$.
 With this substitution, Eq.~(\ref{eq:denominator}) becomes
 \bea
  \frac{x^4}{4}+\frac{1-k_3}{2}x^2+(\frac{1-k_3}{2})^2+k_2^2+x^2-\frac{2x}{r}\Big(\frac{x^4}{4}&+&\frac{1-k_3}{2}x^2+(\frac{1-k_3}{2})^2+k_2^2\Big)\cos
  (\theta+\alpha)\nonumber\\
  &=&\Big((\frac{x^2+1-k_3}{2})^2+k_2^2\Big)\Big(1+\frac{x^2}{(\frac{x^2+1-k_3}{2})^2+k_2^2}-\frac{2x}{r}\cos(\theta+\alpha)\Big)\nonumber\\
  &=&\frac{x^2}{a^2}\Big(1+a^2-2a\cos(\theta+\alpha)\Big),
 \eea
\end{widetext}
where
\bea
 r=\sqrt{\Big(\frac{x^2+1-k_3}{2}\Big)^2+k_2^2}\hspace{20pt}\mathrm{and}\hspace{20pt}a=\frac{x}{r}.
\eea
Then, defining $\gamma=k_1a^2/x^2$, the integral of Eq.~(\ref{eq:integralapp}) becomes
\bea
 &I&=\int\frac{\mathrm{d}^2x}{4\pi}\frac{A}{\Big(k_1 A+\frac{x^2}{a^2}(1+a^2-2a\cos(\theta+\alpha))\Big)^2}\nonumber\\
 &=&
 \int_0^\infty\frac{x\mathrm{d}x}{2\pi}\frac{a^4}{x^4}\int_{0}^{\pi}\frac{A \,\,\mathrm{d}\theta}{\Big(A\gamma+1+a^2-2a\cos(\theta+\alpha)\Big)^2}
\eea after shifting $\theta\rightarrow \theta-\alpha$, and noting
the evenness of the $\theta$ integral. The integral is found in
standard integration tables\cite{Gradshteyn}, and noting that
$(1\pm a)^2+A\gamma\geq0$, we obtain
\bea
 I=\int_0^{\infty}\frac{\mathrm{d}x}{2\pi}\frac{a^4}{x^3}\frac{A\pi(1+a^2)}{(1+a)^3}\Big((1-a)^2+A\gamma\Big)^{-3/2}.
\eea
Since in the limit that $A\rightarrow 0$,
\bea
 \frac{A}{\Big((1+a)^2+A\gamma\Big)^{3/2}}\rightarrow\frac{2}{\gamma}\delta(1-a),
\eea
we find that
\be
 I=\int_0^{\infty}\frac{\mathrm{d}x}{4k_1}\frac{x}{\Big(\frac{x^2+1-k_3}{2}\Big)^2+k_2^2}\,\,\delta(a-1).
\ee
Making the further substitution $y=(x^2+1-k_3)/2$,
\bea
 \label{eq:eyewhy}
 I&=&\int_{\stackrel{\underline{1-k_3}}{2}}^{\infty}\,\,\frac{\mathrm{d}y}{2k_1}\frac{1}{y^2+k_2^2}\,\,\delta\,\Big(
 \frac{2y-(1-k_3)}{(y^2+k_2^2)^2}-1\Big)\nonumber\\
 &=&\int_{\stackrel{\underline{1-k_3}}{2}}^{\infty}\,\,\frac{\mathrm{d}y}{4k_1}\frac{y^2+k_2^2}{\Big|k_2^2-y^2+y(1-k_3)\Big|}\nonumber\\
 & &\times\Big(\delta(y-y_{+})+\delta(y-y_{-})\Big),
\eea
where
\be
 \label{eq:ypm}
 y_{\pm}=1\pm\sqrt{k_3-k_2^2}
\ee
are the intersections of the curves $y^2+k_2^2$ and
$2y-(1-k_3)$. It is easily verified that both $y_+$ and $y_-$ are
in the range of integration $\bf{[}\frac{1-k_3}{2},\infty\bf{)}$
($y_-$ just catching the lower bound when $\psi=0$). Then
expanding the denominator of Eq.~(\ref{eq:eyewhy}) using
Eq.~(\ref{eq:ypm}), we find
\begin{widetext}
 \bea
  \Big|k_2^2-y_\pm^2+y_\pm(1-k_3)\Big|=
  2\sqrt{k_3-k_2^2}\,\,\Big|\sqrt{k_3-k_2^2}\pm\frac{1+k_3}{2}\Big|
\eea
so that
\bea
 I&=&\frac{1}{2k_1}\frac{1}{2\sqrt{k_3-k_2^2}}\Big(\frac{1+k_3+2\sqrt{k_3-k_2^2}}{1+k_3+2\sqrt{k_3-k_2^2}}+\frac{1+k_3-2\sqrt{k_3-k_2^2}}{
 1+k_3-2\sqrt{k_3-k_2^2}}\Big)\nonumber\\
 &=&\frac{1}{2k_1\sqrt{k_3-k_2^2}}
\eea
\end{widetext}
%%%%%%%%%%%%%%%%%%%%%%%%%%%%%%%%%%%%%%%%%%%%%%%%%%%%%%%%%%%%%%%%%%%%%%%%%%%%%%%%%%%%%%%%%%%%%%%%%%%%%%%%%%%%%%%%%%%%%%

%%%%%%%%%%%%%%%%%%%%%%%%%%%%%%%%%%%%%%%%%%%%%%%%%%%%%%%%%%%%%%%%%%%%%%%%%%%%%%%%%%%%%%%%%%%%%%%%%%%%%%%%%%%%%%%%%%%%%%

\begin{thebibliography}{99}

\bibitem{har01} D.J. Van Harlingen, Rev.\ Mod.\ Phys.\ {\bf 67} 515 (1995).
\bibitem{lee02} P.A. Lee, Science {\bf 277}, 50 (1997).
\bibitem{alt01} A. Altland, B.D. Simons and M.R. Zirnbauer, Phys.\ Reports {\bf 359} 283 (2002)
\bibitem{ore01} J. Orenstein and A.J. Millis, Science {\bf 288}, 468 (2000).

\bibitem{gor01} L.P. Gor'kov and P.A. Kalugin, Pis'ma Zh. ksp. Teor. Fiz. {\bf 41}, 208 (1985) [JETP Lett. {\bf 41}, 253 (1985)]
\bibitem{lee01} P.A. Lee, Phys.\ Rev.\ Lett.\ {\bf 71}, 1887 (1993).
\bibitem{hir01} P.J. Hirschfeld, W.O. Putikka and D.J. Scalapino, Phys.\ Rev.\ Lett.\ {\bf 71}, 3705 (1993).
\bibitem{hir02} P.J. Hirschfeld, W.O. Putikka and D.J. Scalapino, Phys.\ Rev.\ B {\bf 50}, 10250 (1994).
\bibitem{hir03} P.J. Hirschfeld and W.O. Putikka, Phys.\ Rev.\ Lett.\ {\bf 77}, 3909 (1996).
\bibitem{graf01} M.J. Graf, S-K. Yip., J.A. Sauls and D. Rainier, Phys.\ Rev.\ B {\bf 53} 15147 (1996).
\bibitem{sen01} T. Senthil, M.P.A. Fisher, L. Balents and C. Nayak, Phys.\ Rev.\ Lett.\ {\bf 81}, 4704 (1998).
\bibitem{dur01} A.C. Durst and P.A. Lee, Phys.\ Rev.\ B {\bf 62} 1270 (2000).

\bibitem{tai01} L. Taillefer, B. Lussier, R. Gagnon, K. Behnia and H. Aubin, Phys.\ Rev.\ Lett.\ {\bf 79}, 483 (1997).
\bibitem{chi01} M. Chiao, R.W. Hill, C. Lupien, B. Popi$\acute{c}$, R. Gagnon and L. Taillefer, Phys.\ Rev.\ Lett.\ {\bf 82} 2943 (1999)
\bibitem{chi02} M. Chiao, R.W. Hill, C. Lupien, L. Taillefer, P. Lambert, R. Gagnon and P. Fournier, Phys.\ Rev.\ B {\bf 62} 3554 (2000)
\bibitem{nak01} S. Nakamae, K. Behnia, L. Balicas, F. Rullier-Albenque, H. Berger and T. Tamegai, Phys. Rev. B {\bf 63} 184509 (2001)
\bibitem{pro01} C. Proust, E. Boaknin, R.W. Hill, L. Taillefer and A.P. Mackenzie, Phys.\ Rev.\ Lett.\ {\bf 89}, 147003 (2002).
\bibitem{sut01} M. Sutherland, D.G. Hawthorn, R.W. Hill, F. Ronning, S. Wakimoto, H. Zhang, C. Proust, E. Boaknin, C. Lupien, L. Taillefer, R.X. Liang, D.A. Bonn, W.N. Hardy, R. Gagnon, N.E. Hussey, T. Kimura, M. Nohara and H. Tagaki, Phys. Rev. B {\bf 67} 174520 (2003)
\bibitem{hil01} R.W. Hill, C. Lupien, M. Sutherland, E. Boaknin, D.G. Hawthorn, C. Proust, F. Ronning, L. Taillefer, R. Liang, D.A. Bonn and W.N. Hardy, Phys.\ Rev.\ Lett.\ {\bf 92}, 027001 (2004).
\bibitem{sun01} X.F. Sun, K. Segawa and Y. Ando, Phys.\ Rev.\ Lett.\ {\bf 93}, 107001 (2004).
\bibitem{sut02} M. Sutherland, S.Y. Li, D.G. Hawthorn, R.W. Hill, F. Ronning, M.A. Tanatar, J. Paglione, H. Zhang, L. Taillefer, J. DeBenedictis, R. Liang, D.A. Bonn and W.N. Hardy, Phys.\ Rev.\ Lett.\ {\bf 94}, 147004 (2005).
\bibitem{haw01} D.G. Hawthorn, S.Y. Li, M. Sutherland, E. Boaknin, R.W. Hill, C. Proust, F. Ronning, M.A. Tanatar, J.P. Paglione, L. Taillefer, D. Peets, R.X. Liang, D.A. Bonn, W.N. Hardy and N.N. Kolesnikov, Phys. Rev. B {\bf 75} 104518 (2007)
\bibitem{sun02} X.F. Sun, S. Ono, X. Zhao, Z.Q. Pang, Y. Abe and Y. Ando, Phys.\ Rev.\ B {\bf 77} 094515 (2008).

\bibitem{kiv01} S.A. Kivelson, I.P. Bindloss, E. Fradkin, V. Oganesyan, J.M. Tranquada, A. Kapitulnik and C. Howald, Rev.\ Mod.\ Phys.\ {\bf 75}, 1201 (2003) (and references within)
\bibitem{pod01} D. Podolsky, E. Demler, K. Damle and B.I. Halperin, Phys. Rev. B {\bf 67} 094514 (2003)
\bibitem{li01} J.X. Li, C.Q. Wu and D.H. Lee, Phys. Rev. B {\bf 74} 184515 (2006)
\bibitem{che01} C.T. Chen, A.D. Beyer and N.C. Yeh, Solid State Communications {\bf 143} 447 (2007)
\bibitem{seo01} K.J. Seo, H.D. Chen, and J.P. Hu, Phys. Rev. B {\bf 76} 020511 (R) (2007)


\bibitem{hof01} J.E. Hoffmann, E.W. Hudson, K.M. Lang, V. Madhavan, H. Eisaki, S. Uchida, and J.C. Davis, Science {\bf 295} 466 (2002).
\bibitem{hof02} J.E. Hoffmann, K. McElroy, D.H. Lee, K.M. Lang, H. Eisaki, S. Uchida and J.C. Davis, Science {\bf 297}, 1148 (2002).
\bibitem{how01} C. Howald, H. Eisaki, N. Kaneko, M. Greven and A. Kapitulnik, Phys. Rev. B {\bf 67} 014533 (2003).
\bibitem{ver01} M. Vershinin, S. Misra, S. Ono, Y. Abe, Y. Ando and A. Yazdani, Science {\bf 303}, 1995 (2004).
\bibitem{mce01} K. McElroy, D.H. Lee, J.E. Hoffman, K.M. Lang, J. Lee, E.W. Hudson, H. Eisaki, S. Uchida, and J.C. Davis, Phys.\ Rev.\ Lett.\ {\bf 94}, 197005 (2005).
\bibitem{han01} T. Hanaguri, C. Lupien, Y. Kohsaka, D.H. Lee, M. Azuma, M. Takano, H. Takagi and J.C. Davis, Nature {\bf 430} 1001 (2004).
\bibitem{mis01} S. Misra, M. Vershenin, P. Phillips and A. Yazdani, Phys.\ Rev.\ B {\bf 70} 220503(R) (2004).
\bibitem{mce02} K. McElroy, D.H. Lee, J.E. Hoffman, K.M. Lang, J. Lee, E.W. Hudson, H. Eisaki, S. Uchida and J.C. Davis, Phys.\ Rev.\ Lett.\ {\bf 94}, 197005 (2005).
\bibitem{koh01} Y. Kohsaka, C. Taylor, K. Fujita, A. Schmidt, C. Lupien, T. Hanaguri, M. Azuma, M. Takano, H. Eisaki, H. Takagi, S. Uchida and J.C. Davis, Science {\bf 315}, 1380 (2007).
\bibitem{boy01} M.C. Boyer, W.D. Wise, K. Chatterjee, M. Yi, T. Kondo, T. Takeuchi, H. Ikuta and E.W. Hudson, Nature Physics {\bf 3}, 802 (2007).
\bibitem{han02} T. Hanguri, Y. Kohsaka, J.C. Davis, C. Lupien, I. Yamada, M. Azuma, M. Takano, K. Ohishi, M. Ono and H. Takagi, Nature Physics {\bf 3}, 865 (2007).
\bibitem{pas01} A.N. Pasupathy, A. Pushp, K.K. Gomes, C.V. Parker, J. Wen, Z. Xu, G. Gu, S. Ono, Y. Ando and A. Yazdani, Science {\bf 320}, 196 (2008).
\bibitem{wis01} W.D. Wise, M.C. Boyer, K. Chatterjee, T. Kondo, T. Takeuchi, H. Ikuta, Y. Wang and E.W. Hudson, Nature Physics {\bf 4}, 696 (2008).
\bibitem{koh02} Y. Kohsaka, C. Taylor, P. Wahi, A. Schmidt, J. Lee, K. Fujita, J.W. Allredge, K. McElroy, J. Lee, H. Eisaki, S. Uchida, D.H. Lee and J.C. Davis, Nature {\bf 454}, 1072 (2008).

\bibitem{ber01} E. Berg, C.C. Chen and S.A. Kivelson, Phys.\ Rev.\ Lett.\ {\bf 100} 027003 (2008)
\bibitem{par01} K. Park and S. Sachdev, Phys.\ Rev.\ B {\bf 64} 184510 (2001)
\bibitem{gra01} M. Granath, V. Oganesyan, S. A. Kivelson, E. Fradkin, and V. J. Emery, Phys.\ Rev.\ Lett.\ {\bf 86}, 167011 (2001)
\bibitem{voj01} M. Vojta, Y. Zhang and S. Sachdev, Phys\. Rev.\ B {\bf 62} 6721 (2000)

\bibitem{dur02} A. C. Durst and S. Sachdev, arXiv:0810.3914 (2008)

\bibitem{hus01} N.E. Hussey, Advances in Physics {\bf 51}, 1685 (2002).
\bibitem{tak01} J. Takeya, Y. Ando, S. Komiya, an X. F. Sun, Phys.\ Rev.\ Lett.\ {\bf 88}, 077001 (2002).
\bibitem{sun05} X. F. Sun, S. Komiya, J. Takeya, and Y. Ando, Phys.\ Rev.\ Lett.\ {\bf 90}, 117004 (2003).
\bibitem{and01} Y. Ando, S. Ono, X.F. Sun, J. Takeya, F.F. Balakirev, J.B. Betts and G.S. Boebinger, Phys.\ Rev.\ Lett.\ {\bf 92}, 247004 (2004).
\bibitem{sun03} X.F. Sun, K. Segawa and Y. Ando, Phys.\ Rev.\ B {\bf 72}, 100502 (2005).
\bibitem{sun04} X.F. Sun, S. Ono, Y. Abe, S. Komiya, K. Segawa and Y. Ando, Phys.\ Rev.\ Lett.\ {\bf 96}, 017008 (2006).
\bibitem{haw02} D.G. Hawthorn, R.W. Hill, C. Proust, F. Ronning, M. Sutherland, E. Boaknin, C. Lupien, M.A. Tanatar, J. Paglione, S. Wakimoto, H. Zhang, L. Taillefer, T. Kimura, M. Nohara and N.E. Hussey, Phys.\ Rev.\ Lett.\ {\bf 90}, 197004 (2003).

\bibitem{gus01} V. P. Gusynin and V. A. Miransky, Eur.\ Phys.\ J.\ B {\bf 37}, 363 (2004).
\bibitem{ander01} B. M. Andersen and P. J. Hirschfeld, Phys.\ Rev.\ Lett.\ {\bf 100}, 257003 (2008).

\bibitem{mah01} G. D. Mahan, {\it Many-Particle Physics} (Plenum Press, New York, 1981).
\bibitem{Fetter} A. L. Fetter and J. D. Walecka, {\it Quantum Theory of Many-Particle Systems} (McGraw Hill, Boston, 1971).
\bibitem{Gradshteyn} I. S. Gradshteyn and I. M. Ryzhik, {\it Table of Integrals, Series and Products} (Academic Press, San Diego, 1994).

\end{thebibliography}
\end{document}